\documentclass[aps,twocolumn,superscriptaddress]{revtex4-1}
\usepackage{amssymb, amsmath, bm}
\usepackage{graphicx,onlyamsmath}

\usepackage{natbib}
\usepackage{xcolor}
\usepackage[normalem]{ulem}  
\usepackage{makecell} 

\begin{document}

\title{Higher-order topological insulator in cubic semiconductor quantum wells}

\author{Sergey S.~Krishtopenko}
\email[]{sergey.krishtopenko@gmail.com}
\affiliation{CENTERA Laboratories, Institute of High Pressure Physics, Polish Academy of Sciences, PL-01-142 Warsaw, Poland}
\affiliation{Laboratoire Charles Coulomb (L2C), UMR 5221 CNRS-Universit\'{e} de Montpellier, F- 34095 Montpellier, France}
\date{\today}

\begin{abstract}
The search for exotic new topological states of matter in widely accessible materials, for which the manufacturing process is mastered, is one of the major challenges of the current topological physics. Here we predict higher order topological insulator state in quantum wells based on the most common semiconducting materials. By successively deriving the bulk and boundary Hamiltonians, we theoretically prove the existence of topological corner states due to cubic symmetry in quantum wells with double band inversion. We show that the appearance of corner states does not depend solely on the crystallographic orientation of the meeting edges, but also on the growth orientation of the quantum well. Our theoretical results significantly extend the application potential of topological quantum wells based on IV, II-VI and III-V semiconductors with diamond or zinc-blende structures.
\end{abstract}

\maketitle

\section{Introduction}
Since the discovery of $\mathbb{Z}_2$ topological insulators (TIs)~\cite{w1,w2}, topological phases and materials have been extensively explored in two- (2D) and three-dimensional (3D) systems~\cite{w3,w4,w5}. The recent classification~\cite{w6,w7,w8} of TIs with given crystalline symmetry has led to the discovery of a new type of topological phases, the higher-order topological insulators (HOTIs)~\cite{w9,w10,w11,w12,w13,w14,w15}. The 3D HOTIs are gapped in the bulk and on all surfaces, but they have one-dimensional (1D) gapless modes along ``hinges'', where two surfaces meet. These hinge states were experimentally observed in bismuth~\cite{w16}, Bi$_4$Br$_4$~\cite{w16d} and WTe$_2$ crystals~\cite{w16b,w16c} and theoretically predicted for strained SnTe~\cite{w12}, transition metal dichalcogenides $X{\mathrm{Te}}_{2}$~\cite{w17} and antiperovskites~\cite{w18}.

In 2D HOTIs, pioneering works~\cite{w9,w10,w13} suggest the presence of zero-dimensional (0D) corner states inside the insulating edge and bulk band-gap of certain materials. Recently, Peterson~\emph{et~ al.}~\cite{w15b} have shown that 0D corner states may reside either in the bulk band-gap or fully within the bulk bands of a HOTI, depending on the material's details. HOTIs that fall into the latter case do not host 0D corner states within their bulk band-gap and, as such, cannot be distinguished from trivial insulators by their spectrum alone. Nevertheless, even in this case the higher-order topology can be still identified via a fractional corner anomaly~\cite{w15b,w15c}.

So far, a large part of the experimental study of 2D HOTIs has been performed in engineered metamaterials~\cite{w22,w23,w24,w25,w26,w27,w28,w30,w31,w32,w33}, while only a few candidates have been theoretically predicted in solids, including black phosphorene~\cite{w19}, graphdiyne~\cite{w20}, bismuthene~\cite{w20b} and twisted bilayer graphene at certain angles~\cite{w21}. Although twisted bilayer graphene can be indeed a realistic candidate to probe 2D HOTI state experimentally, it is still highly desirable to identify controllable and widely accessible higher-order topological materials.

Nowadays, many technologically important semiconductors with the most developed molecular-beam-epitaxy growth hold a cubic crystal structure, including the diamond structure for the group-IV elements, and the zinc-blend structure for the III-V and II-VI compounds. For instance, the time-reversal-invariant 2D TI state, also known as quantum spin Hall insulator (QSHI) -- was first discovered in the cubic semiconductor QWs with an inverted band structure: HgTe/CdHgTe QWs~\cite{w2,w34} and broken-gap InAs/GaSb QW bilayers~\cite{w35,w36}. Later, InN/GaN QWs~\cite{w37}, Ge/GaAs QWs~\cite{w38} and InAsBi/AlSb QWs~\cite{w39} were also predicted to be 2D~TIs. Such a list can be obviously extended by including variety of type-II broken-gap QW heterostructures~\cite{w40,w41,w42,w43} (similar to the InAs/GaSb QWs) on the basis of III-V semiconductors and their alloys.

This work shows that in addition to ``conventional'' first-order 2D TI state, cubic IV, III-V and II-VI semiconductors are also promising for the implementation of time-reversal-invariant 2D HOTI.
Starting from realistic multi-band \textbf{k$\cdot$p} Hamiltonian~\cite{w46}, we directly derive an effective 2D low-energy Hamiltonian preserving the cubic symmetry of the semiconductors. Then, by applying open boundary conditions, we obtain an effective 1D Hamiltonian for the edge states and demonstrate the existence of the corner states in the QWs with double band inversion. As two prototype 2D systems, we consider three-layer InAs/GaInSb QWs~\cite{w45} and double HgTe/CdHgTe QWs~\cite{w44} grown along ($0mn$) crystallographic orientations (where $m$ and $n$ are integers).

\section{Results}
\subsection{Insulating state with double band inversion}
\begin{figure}
\includegraphics [width=1.0\columnwidth, keepaspectratio] {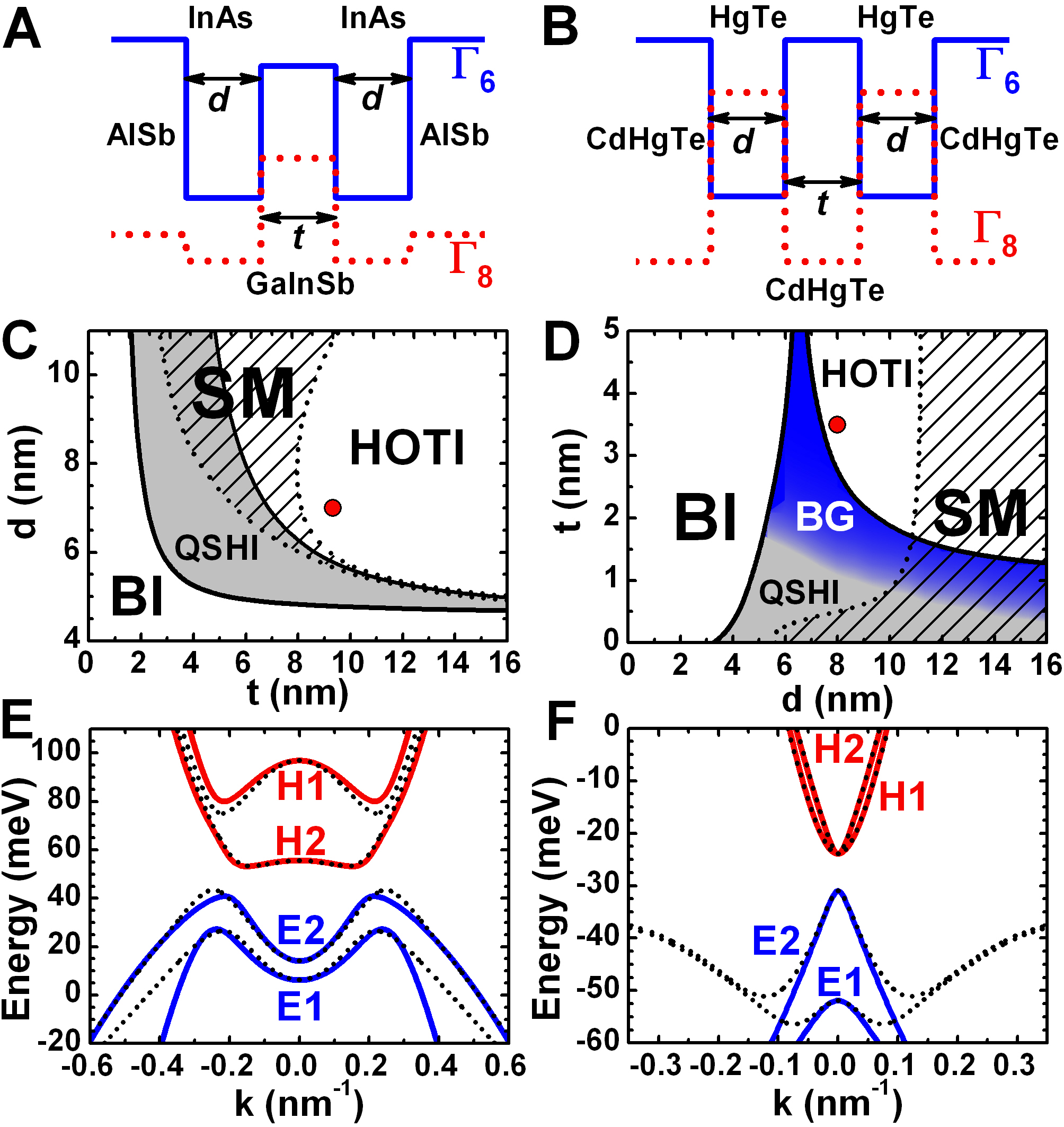} 
\caption{\label{Fig:1} \textbf{Band structure of three-layer InAs/GaInSb and double HgTe/CdHgTe QWs.}
(\textbf{A}) Schematic representation of symmetrical three-layer InAs/Ga$_{0.65}$In$_{0.35}$Sb QW confined by outer AlSb barriers~\cite{w45}. Here, $d$ and $t$ are the thicknesses of InAs and Ga$_{0.65}$In$_{0.35}$Sb layers, respectively. The QW is supposed to be grown on (001) GaSb buffer. (\textbf{B}) Schematic representation of double HgTe/Cd$_{0.7}$Hg$_{0.3}$Te QW. Here, $d$ is the thickness of HgTe layers and $t$ is the middle CdHgTe barrier thickness. The double QW is assumed to be grown on (001) CdTe buffer. The Hg content in all the barriers is chosen to be $0.3$~\cite{w44}.
(\textbf{C,D}) The phase diagrams for different $d$ and $t$. The left-hand and right-hand solid curves correspond to the crossing between \emph{E}1--\emph{H}1 subbands and \emph{E}2--\emph{H}2 subbands, respectively. These curves divide the plane into three parts with trivial band ordering corresponding to band insulator (BI, see the left-hand white region), single band inversion (grey and blue regions) and double-band inversion (right-hand white region). The striped region defines a semimetal (SM) phase with vanishing \emph{indirect} band-gap~\cite{w44,w45}.
(\textbf{E,F}) Band structure calculated on the basis of effective 2D low-energy Hamiltonian for the QWs with the layer thicknesses marked by the red symbols in (B) and (C) panels. The blue and red curves represent band dispersion of electron-like and hole-like subbands, respectively. The wave vector is oriented along (100) crystallographic direction. The dotted curves represent the calculations based on realistic multi-band \textbf{k$\cdot$p} Hamiltonian~\cite{w46}.}
\end{figure}

Let us first explore the possibility for double band inversion in the prototype QWs (see Fig.~\ref{Fig:1}). Since these QWs can be considered as two tunnel-coupled HgTe QWs~\cite{w2,w34} or InAs/GaInSb QW bilayers~\cite{w35,w36}, each of which features a single band inversion, the appearance of double band inversion is not surprising for such multi-layer systems. In Figs~\ref{Fig:1}C and \ref{Fig:1}D, we provide a phase diagram for three-layer InAs/Ga$_{0.65}$In$_{0.35}$Sb and double HgTe/Cd$_{0.7}$Hg$_{0.3}$Te QWs with different layer thicknesses. Both QWs are supposed to be grown on the (001) crystallographic plane.

In the diagrams, the left-hand black solid curve describing the crossing between the first electron-like (\emph{E}1) and hole-like (\emph{H}1) subbands divides the $d$-$t$ plane into a white region, corresponding to band insulator (BI) with trivial band ordering, and a grey region of QSHI with inverted band structure. If the middle barrier is thick enough, in addition to QSHI, the double HgTe/CdHgTe QWs also hold a specific state with a band structure similar to the one of bilayer graphene (BG) (see the blue region in Fig.~\ref{Fig:1}D). A detailed discussion of the ``bilayer graphene'' state can be found in~\cite{w44}.

Further increasing of $d$ and $t$ results in the band crossing between the second electron-like (\emph{E}2) and hole-like (\emph{H}2) subbands, which is shown by the right-hand black solid curve in the diagram. This curve, in its turn, separates the grey and blue regions with single band inversion from the right-hand white region corresponding to the double band inversion, when
two electron-like \emph{E}1 and \emph{E}2 levels lie below two hole-like \emph{H}1 and \emph{H}2 subbands.

Finally, at certain $d$ and $t$ values, corresponding to the striped region, the so-called semimetal (SM) phase is implemented. The semimetal phase is characterized by a vanishing \emph{indirect} band-gap when the side maxima of the valence subband exceed in energy the conduction subband bottom~\cite{w44,w45}. Thus, by varying the layer thicknesses in the prototype QWs, one can indeed realize band insulator, QSHI (or ``bilayer graphene'' state), semimetal phase and the insulator state with a double band inversion.
In this work, we identify a double-band-inversion insulator state 2D HOTI with the corner states arising due to cubic symmetry of II-VI and III-V semiconductors.

To describe double band inversion at the $\Gamma$ point of the Brillouin zone, we derive an effective 2D low-energy Hamiltonian taking into account \emph{E}1, \emph{E}2, \emph{H}1 and \emph{H}2 subbands and preserving the cubic symmetry of the prototype QWs. Starting from a realistic multi-band \textbf{k$\cdot$p} Hamiltonian for ($0mn$)-oriented cubic semiconductor QWs~\cite{w46} and following  expansion procedure described in the Supplementary Materials, in the basis $|E1{+}\rangle$, $|H1{+}\rangle$, $|H2{-}\rangle$, $|E2{-}\rangle$, $|E1{-}\rangle$, $|H1{-}\rangle$, $|H2{+}\rangle$, $|E2{+}\rangle$, the effective 2D Hamiltonian in the vicinity of the $\Gamma$ point has the form:
\begin{equation}
\label{eq:1}
H_{2D}(k_x,k_y)=\begin{pmatrix}
H_{4\times4}(k_x,k_y) & 0 \\ 0 & H_{4\times4}^{*}(-k_x,-k_y)\end{pmatrix},
\end{equation}
where the asterisk denotes complex conjugation. The diagonal blocks of $H_{2D}(k_x,k_y)$ are split into isotropic and anisotropic parts:
\begin{equation}
\label{eq:2}
H_{4\times4}(k_x,k_y)=H_{4\times4}^{(i)}(k_x,k_y)+H_{4\times4}^{(a)}(k_x,k_y).
\end{equation}
The isotropic part $H_{4\times4}^{(i)}(k_x,k_y)$ is written as~\cite{w47,w48}:
\begin{equation*}
\label{eq:3}
H_{4\times4}^{(i)}(k_x,k_y)=\begin{pmatrix}
\epsilon_{E1} & -A_{1}k_{+} & R_{1}^{(i)}k_{-}^2 & S_{0}k_{-}\\
-A_{1}k_{-} & \epsilon_{H1} & 0 & R_{2}^{(i)}k_{-}^2\\
R_{1}^{(i)}k_{+}^2 & 0 & \epsilon_{H2}  & A_{2}k_{+}\\
S_{0}k_{+} & R_{2}^{(i)}k_{+}^2 & A_{2}k_{-} & \epsilon_{E2} \end{pmatrix},
\end{equation*}
where
\begin{eqnarray}
\label{eq:4}
\epsilon_{E1}(k_x,k_y)=C_1+M_1-(D_{1}+B_{1})(k_x^2+k_y^2),~\nonumber\\
\epsilon_{H1}(k_x,k_y)=C_1-M_1-(D_{1}-B_{1})(k_x^2+k_y^2),~\nonumber\\
\epsilon_{E2}(k_x,k_y)=C_2+M_2-(D_{2}+B_{2})(k_x^2+k_y^2),~\nonumber\\
\epsilon_{H2}(k_x,k_y)=C_2-M_2-(D_{2}-B_{2})(k_x^2+k_y^2).~
\end{eqnarray}
Here, $k_{\pm}=k_x+ik_y$, $k_x$ and $k_y$ are the momentum components in the QW plane, and $C_{1,2}$, $M_{1,2}$, $A_{1,2}$, $B_{1,2}$, $D_{1,2}$, $S_0$ and $R_{1,2}^{(i)}$ are \emph{isotropic} structure parameters being defined by the QW geometry, the growth orientation and the materials. The Hamiltonian $H_{2D}(k_x,k_y)$ has a block-diagonal form because we keep the inversion symmetry (see Supplementary Materials) by neglecting the terms resulting from the anisotropy of chemical bonds at the QW interfaces~\cite{w49} and the bulk inversion asymmetry of the unit cell of zinc-blende semiconductors~\cite{w50}. The latter is absent for diamond-like semiconductors.

The most important quantities in $H_{2D}(k_x,k_y)$ are two mass parameters $M_1$ and $M_2$ describing the band inversion between \emph{E}1--\emph{H}1 subbands and \emph{E}2--\emph{H}2 subbands, respectively. The trivial band insulator corresponds to positive values of $M_1$ and $M_2$. The QSHI and ``bilayer graphene'' states arise if $M_1<0$ and $M_2>0$, and the difference between these states is defined by the gap between \emph{H}1 and \emph{H}2 subbands, which is zero in the case of ``bilayer graphene'' state~\cite{w44}. The insulator state with double band inversion is defined by the negative values of $M_1$ and $M_2$. We note that since the semimetal phase represented by the striped areas in the diagrams is formed by non-local overlapping of the valence and conduction subbands, it cannot be described within the low-energy Hamiltonian for the small values of $k_x$ and $k_y$.

The isotropic term $H_{4\times 4}^{(i)}(k_x,k_y)$ preserves the rotational symmetry in the QW plane, therefore it is independent of the orientation of $x$ and $y$ axis. In contrast, the form of the anisotropic term $H_{4\times4}^{(a)}(k_x,k_y)$ in Eq.~(\ref{eq:2}), resulting from the cubic symmetry of diamond and zinc-blende semiconductors, depend not only on the QW growth orientation but also on the orientation of $x$ and $y$ axis (see Supplementary Materials). For ($0mn$)-oriented QWs, $H_{4\times4}^{(a)}(k_x,k_y)$ has the form
\begin{multline*}
{H}_{4\times4}^{(a)}({k}_x,{k}_y)= \\
-\begin{pmatrix}
0 & 0 & R_{1}^{(a)}e^{i4\varphi}{k}_{+}^2 & 0\\
0 & 0 & 0 & R_{2}^{(a)}e^{i4\varphi}{k}_{+}^2\\
R_{1}^{(a)}e^{-i4\varphi}{k}_{-}^2 & 0 & 0  & 0\\
0 & R_{2}^{(a)}e^{-i4\varphi}{k}_{-}^2 & 0 & 0 \end{pmatrix} \\
-\left({k}_{y}\cos{\varphi}+{k}_{x}\sin{\varphi}\right)^{2}\sin^{2}{2\theta}\times \\
\times\begin{pmatrix}
0 & 0 & R_{1}^{(a)}e^{i2\varphi} & 0\\
0 & 0 & 0 & R_{2}^{(a)}e^{i2\varphi}\\
R_{1}^{(a)}e^{-i2\varphi} & 0 & 0  & 0\\
0 & R_{2}^{(a)}e^{-i2\varphi} & 0 & 0 \end{pmatrix}+
\end{multline*}
\begin{multline}
\label{eq:5}
+\sin{2\theta}\begin{pmatrix}
0 & 0 & \tilde{R}_{1}^{(a)}e^{i2\varphi} & 0\\
0 & 0 & 0 & \tilde{R}_{2}^{(a)}e^{i2\varphi}\\
\tilde{R}_{1}^{(a)}e^{-i2\varphi} & 0 & 0  & 0\\
0 & \tilde{R}_{2}^{(a)}e^{-i2\varphi} & 0 & 0 \end{pmatrix},
\end{multline}
where $R_{1,2}^{(a)}$ and $\tilde{R}_{1,2}^{(a)}$ are \emph{cubic} structure parameters, which depend on the QW geometry and materials. In Eq.~(\ref{eq:5}), $\theta=\arctan(m/n)$ is the angle defining the QW growth orientation, while the angle $\varphi$ is the angle between the $x$ axis and the ($001$) crystallographic direction.

As clear from above, depending on the structure parameters, the effective 2D Hamiltonian in Eq.~(\ref{eq:1}) describes QSHI, ``bilayer-graphene'' state, trivial band insulator or insulator with double band inversion. Note that thirteen parameters involved in $H_{2D}(k_x,k_y)$ cannot take arbitrary values.
Since their calculation is based on the wave-functions of the multi-band \textbf{k$\cdot$p} Hamiltonian at $k_x=k_y=0$ (for details, see Supplementary Materials), the set of parameters corresponding to specific topological state is determined by the thicknesses and materials of the QW layers, its growth orientation, and the buffer on which the QW is grown. The latter is crucial for taking into account the effect of lattice-mismatch strain on the band structure in the QW. Thus, all of the structure parameters are the functions of $d$, $t$, $m$, $n$ and the QW layer and buffer materials. Further, we perform the calculations for two sets of structure parameters involved in $H_{2D}(k_x,k_y)$, which correspond to the prototype QWs with the layer thicknesses marked by the red symbols in Figs~\ref{Fig:1}(C) and \ref{Fig:1}(D). Figures~\ref{Fig:1}E and \ref{Fig:1}F compare the band structure calculations based on realistic multi-band \textbf{k$\cdot$p} Hamiltonian~\cite{w46} and $H_{2D}(k_x,k_y)$ in Eq.~(\ref{eq:1}). The isotropic and cubic structure parameters of $H_{2D}(k_x,k_y)$ for the insulator state with double band inversion are given in the Supplementary Materials.

\subsection{Anisotropic edge states}
Let us now analyze the edge states arising in the insulator with double band inversion. Since $H_{2D}(k_x,k_y)$ in Eq.~(\ref{eq:1}) has block-diagonal form,  we further focus on the upper block $H_{4\times4}(k_x,k_y)$ only, while the calculations for the lower block $H_{4\times4}^{*}(-k_x,-k_y)$ can be performed in the similar manner.

To derive an effective 1D low-energy edge Hamiltonian, we split $H_{2D}(k_x,k_y)$ into two parts so that the first part represents two independent BHZ-like models~\cite{w2} with $M_1<0$ and $M_2<0$ for the pairs of \emph{E}1--\emph{H}1 subbands and \emph{E}2--\emph{H}2 subbands, while the second part includes the rest isotropic and cubic terms describing the inter-pairs mixing. Then, assuming open-boundary conditions in a semi-infinite plane $y>0$, we solve the eigenvalue problem for the independent BHZ-like blocks to find the edge wave-functions at $k_x=0$. In this case, the edge orientation represented by the $x$ axis is defined by the angle $\varphi$ measured from ($100$) crystallographic direction. Finally, we construct a low-energy edge Hamiltonian by projecting $H_{4\times4}(k_x,k_y)$ onto the obtained set of the basis edge functions (see Supplementary Materials).

The projection of two independent BHZ-like blocks~\cite{w2} with non-zero $k_x$ leads to
\begin{multline*}
\label{eq:6}
H_{1D}^{(0)}(k_{x})=~~~~~~ \\ \begin{pmatrix}
C_1-\dfrac{M_1D_1}{B_1}-\dfrac{2A_1\eta_1}{1+\eta_1^2}k_x &  0 \\
0 & C_2-\dfrac{M_2D_2}{B_2}+\dfrac{2A_2\eta_2}{1+\eta_2^2}k_x
\end{pmatrix},
\end{multline*}
where $\eta_{n}^2=(B_{n}+D_{n})/(B_{n}-D_{n})$ with $n=1$ and $2$ corresponding to the pairs of \emph{E}1-\emph{H}1 and \emph{E}2-\emph{H}2 subbands, respectively. One can see that $H_{1D}^{(0)}(k_{x})$ describes the linear edge dispersion due to the inversion of the subband pairs $|E1,{+}\rangle$--$|H1,{+}\rangle$ and $|E2,{-}\rangle$--$|H2,{-}\rangle$ in the absence of their mixing~\cite{w51}. The two energy branches cross at $k_x=k_c$:
\begin{equation}
\label{eq:7}
k_c=\dfrac{C_1-C_2+\dfrac{M_2D_2}{B_2}-\dfrac{M_1D_1}{B_1}}{\dfrac{2A_1\eta_1}{1+\eta_1^2}+\dfrac{2A_2\eta_2}{1+\eta_2^2}},
\end{equation}
that allows the representation of $H_{1D}^{(0)}(k_{x})$ in the form of the Hamiltonian of ``tilted'' 1D massless Dirac fermions:
\begin{equation}
\label{eq:8}
H_{1D}^{(0)}(k_{x})=\varepsilon_0+v_0\delta{k}\mathbf{I}_2+v_z\delta{k}\sigma_z,
\end{equation}
where $\delta{k}=k_x-k_c$, $\mathbf{I}_2$ is a $2{\times}2$ identity matrix,  $\sigma_z$ is one of the Pauli matrices, and $\varepsilon_0$ is a constant corresponding to the energy of the crossing point at $k_x=k_c$. In Eq.~(\ref{eq:8}), $v_0$ and $v_z$ are written as
\begin{eqnarray}
\label{eq:9}
v_0=\dfrac{A_1\eta_1}{1+\eta_1^2}-\dfrac{A_2\eta_2}{1+\eta_2^2},\nonumber\\
v_z=\dfrac{A_1\eta_1}{1+\eta_1^2}+\dfrac{A_2\eta_2}{1+\eta_2^2}.
\end{eqnarray}
Note that the crossing of other Kramer's partners occurs at $k_x=-k_c$.

The projection of the rest terms of $H_{4\times4}(k_x,k_y)$ representing the mixing between the pairs $|E1,{+}\rangle$--$|H1,{+}\rangle$ and $|E2,{-}\rangle$--$|H2,{-}\rangle$ results in anti-diagonal mass terms describing the band-gap opening. After straightforward calculations with the details provided in the Supplementary Materials, we finally obtain the low-energy 1D edge Hamiltonian:
\begin{multline}
\label{eq:9}
H_{\mathrm{1D}}(\delta{k},\theta,\varphi)=\varepsilon_0+v_0\delta{k}\mathbf{I}_2+v_z\delta{k}\sigma_z+ \\
+\left(m_y+v_y\delta{k}+\delta_y\delta{k}^2\right)\sigma_y+ \\ +\left(m_x+v_x\delta{k}+\delta_x\delta{k}^2\right)\sigma_x,
\end{multline}
where $v_x$, $v_y$, $m_x$, $m_y$, $\delta_x$, $\delta_y$ include the angle dependence on $\theta$ and $\varphi$:
\begin{align}
\label{eq:10}
\begin{split}
m_x=F_a{\kappa_2}\cos{4\varphi}-F_i{\kappa_2}-F_0\kappa_1+\left(F_i-F_a\cos{4\varphi}\right)k_{c}^2+ \\
+\left[F_0-2{\kappa_1}(F_i+F_a\cos{4\varphi})\right]k_{c}+\tilde{F}_{a}\cos{2\varphi}\sin{2\theta}- \\
-F_{a}\cos{2\varphi}\sin^{2}{2\theta}\left(k_{c}^{2}\sin^{2}{\varphi}
+{\kappa_2}\cos^{2}{\varphi}\right)-\\
-F_{a}\kappa_{1}k_{c}\sin^{2}{2\varphi}\sin^{2}{2\theta},
\end{split}\nonumber\\
\begin{split}
m_y=F_a\sin{4\varphi}\left[k_c^2+2\kappa_1{k_c}-\kappa_2\right]
-\tilde{F}_{a}\sin{2\varphi}\sin{2\theta}+~~~\\
+F_{a}\sin{2\varphi}\sin^{2}{2\theta}
\left(k_{c}^{2}\sin^{2}{\varphi}-\kappa_{1}k_{c}\cos{2\varphi}+{\kappa_2}\cos^{2}{\varphi}\right),
\end{split}\nonumber\\
\begin{split}
v_x=F_{0}+2k_c\left(F_{i}-F_{a}\cos{4\varphi}\right)-2\kappa_1\left(F_{i}+F_{a}\cos{4\varphi}\right)- ~~\\
-F_{a}\sin^{2}{2\theta}\left(2k_{c}\cos{2\varphi}\sin^{2}{\varphi}+\kappa_{1}\sin^{2}{2\varphi}\right),~~
\end{split}\nonumber\\
\begin{split}
v_y=2F_a\sin{4\varphi}\left(\kappa_1+k_c\right)+~~~~~~~~~~~~~~~~~~~~~~~~~~~~~~~~\\
+F_{a}\sin{2\varphi}\sin^{2}{2\theta}\left(2k_{c}\sin^{2}{\varphi}-\kappa_{1}\cos{2\varphi}\right),
\end{split}\nonumber\\
\begin{split}
\delta_x=F_i-F_a\cos{4\varphi}-F_{a}\cos{2\varphi}\sin^{2}{\varphi}\sin^{2}{2\theta},~~~~~~~~~~~~~~~~
\end{split}\nonumber\\
\begin{split}
\delta_y=F_a\sin{4\varphi}+F_{a}\sin{2\varphi}\sin^{2}{\varphi}\sin^{2}{2\theta}.
\end{split}
\end{align}
Here, $\kappa_1$ and $\kappa_2$ are defined by the matrix elements $\left\langle{\partial/\partial{y}}\right\rangle$ and $\left\langle{\partial^2/\partial{y^2}}\right\rangle$, respectively, both calculated by using the basis edge functions at $k_x=0$ (see Supplementary Materials). In Eqs~(\ref{eq:10}), we have also introduced the edge \emph{isotropic} ($F_{i}$, $F_{0}$) and \emph{cubic} ($F_{a}$, $\tilde{F}_{a}$) parameters:
\begin{eqnarray}
\label{eq:11}
F_{i}=\dfrac{R_{1}^{(i)}\eta_2+R_{2}^{(i)}\eta_1}{\sqrt{1+\eta_1^2}\sqrt{1+\eta_2^2}},~~~~
F_{0}=\dfrac{S_0}{\sqrt{1+\eta_1^2}\sqrt{1+\eta_2^2}},~~~~~~~\nonumber\\
F_{a}=\dfrac{R_{1}^{(a)}\eta_2+R_{2}^{(a)}\eta_1}{\sqrt{1+\eta_1^2}\sqrt{1+\eta_2^2}},~~~~
\tilde{F}_{a}=\dfrac{\tilde{R}_{1}^{(a)}\eta_2+\tilde{R}_{2}^{(a)}\eta_1}{\sqrt{1+\eta_1^2}\sqrt{1+\eta_2^2}}.~~~~~~~
\end{eqnarray}
The similar calculations for the block $H_{4\times4}^{*}(-k_x,-k_y)$ in Eq.~(\ref{eq:1}) results in $H_{\mathrm{1D}}^{*}(-k_x-k_c,\theta,\varphi)$. The parameters $v_x$, $v_y$, $m_x$, $m_y$, $\delta_x$, $\delta_y$ as a function of the edge orientation $\varphi$ for the prototype QWs grown along ($001$), ($011$) and ($013$) crystallographic directions are provided in the Supplementary Materials.

\begin{figure}
\includegraphics [width=1.0\columnwidth, keepaspectratio] {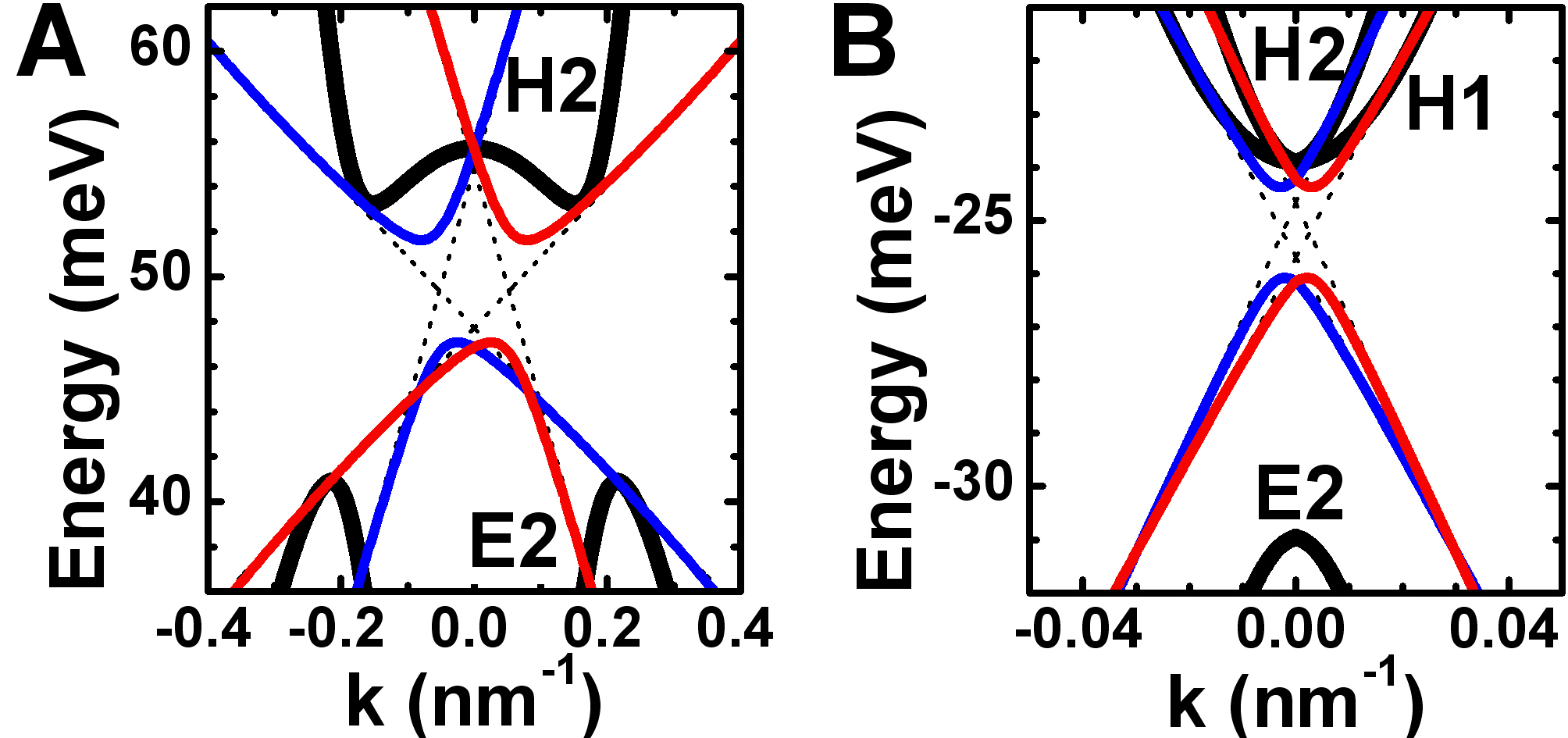} 
\caption{\label{Fig:2} \textbf{Energy dispersion of the bulk and edge states in the prototype QWs with double band inversion.} The (\textbf{A}) and (\textbf{B}) panels presents the calculations based on low-energy 1D and 2D Hamiltonians for the (001)-oriented three-layer InAs/GaInSb and double HgTe/CdHgTe QWs, respectively. The layer thicknesses of the prototype QWs are marked by the red symbols in Figs~\ref{Fig:1}(C) and \ref{Fig:1}(D). Bulk states being described by $H_{2D}(k_x,k_y)$ in Eq.~(\ref{eq:1}) are shown in black. Red and blue curves correspond to different Kramer's partners of the edge states. The dashed curves are the eigenvalues of $H_{1D}^{(0)}(k_{x})$ in Eq.~(\ref{eq:8}). The wave vector and the edge are oriented along (100) crystallographic direction.}
\end{figure}

Figure~\ref{Fig:2} shows the edge state picture of the prototype QWs with double band inversion based on low-energy 1D edge Hamiltonian for the edge oriented along ($100$) crystallographic direction, i.e. $\varphi=0$. One can see that the edge dispersion of each Kramer's partner mimics the dispersion of ``tilted'' 1D massive Dirac fermions. However, an important distinctive property of the edge states shown in Fig.~\ref{Fig:2} is that their gap is described simultaneously by two mass $m_x$ and $m_y$ parameters that prevent the gap vanishing at specific edge orientation (cf. Refs~\cite{w19,w20,w20b,w21}). Indeed, if one neglects the cubic terms of $H_{2D}(k_x,k_y)$, resulting in $F_{a}=\tilde{F}_{a}=0$ in Eqs~(\ref{eq:10}), the mass parameter $m_y$ vanishes, while $m_x$ becomes independent of the edge orientation. Nevertheless, even such complex structure of the edge states yields the corner states in prototype QWs with double band inversion.

\subsection{0D corner states and boundary conditions}
\begin{figure}
\includegraphics [width=1.0\columnwidth, keepaspectratio] {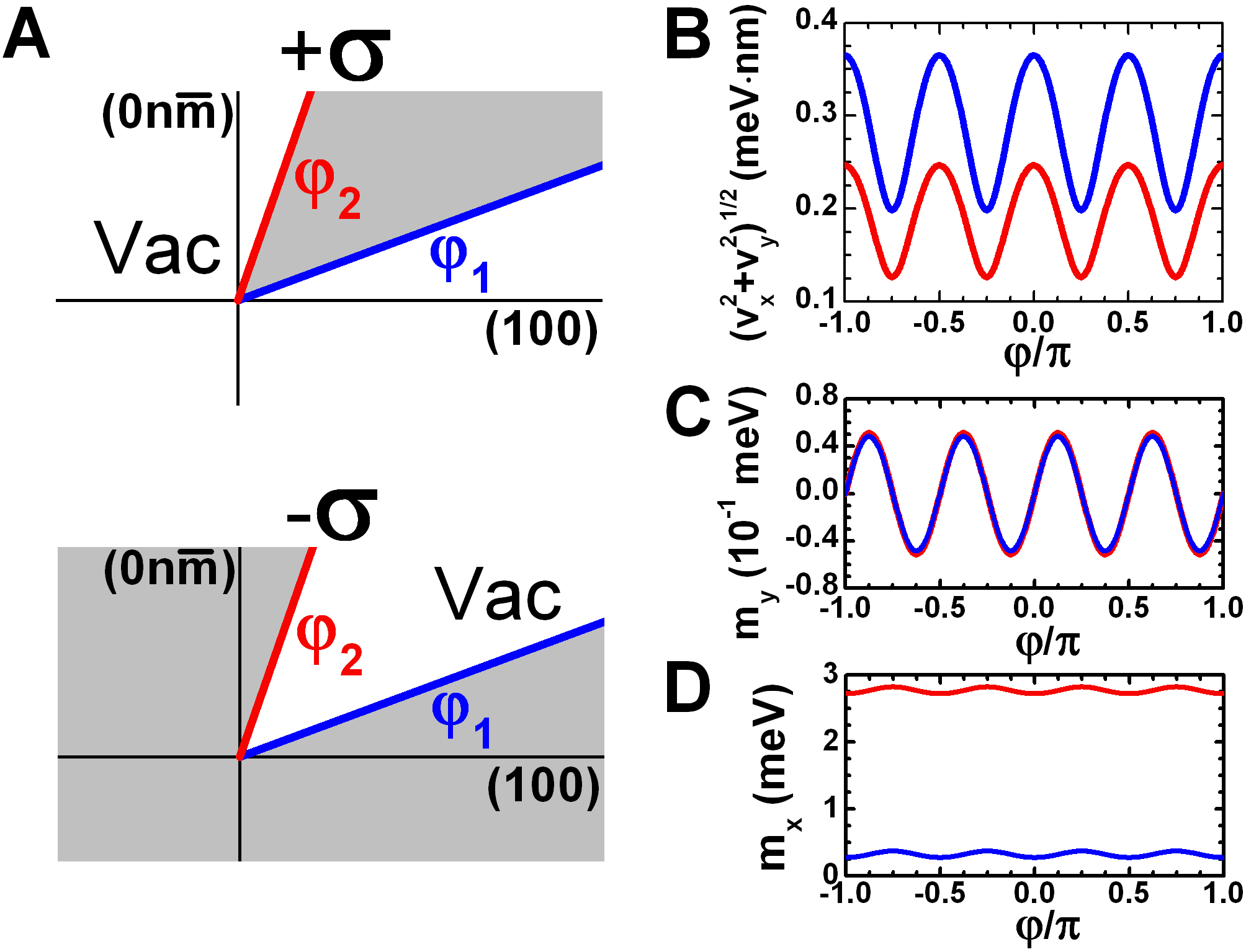} 
\caption{\label{Fig:3} \textbf{Orientation of the corner edge of (0mn)-oriented prototype QWs.}
(\textbf{A}) Schematic of two edges with a common corner with respect to main crystallographic axes in the QW plane. The sample and external vacuum are shown in grey and white, respectively. External and internal angles differ by the parameter $+\sigma$ and $-\sigma$ (see the text). (\textbf{B-D}) Dependence of $m_x$, $m_y$ and $\sqrt{v_x^2+v_y^2}$ on the edge orientation $\varphi$ for the (001)-oriented three-layer InAs/GaInSb (in red) and double HgTe/CdHgTe QW (in blue) with the layer thicknesses marked by the red symbols in Figs~\ref{Fig:1}(C) and \ref{Fig:1}(D). The edge independent parameters ($v_0$, $v_z$) equal to ($36.6$, $68.1$)~meV$\cdot$nm and ($-16.9$, $200.3$)~meV$\cdot$nm for the three-layer InAs/GaInSb and double HgTe/CdHgTe QW, respectively. This means a good approximation of $\sqrt{v_x^2+v_y^2}\ll|v_0|,|v_z|$ for both QWs.}
\end{figure}

To calculate the energy of the corner states, we apply linear approximation of the effective low-energy 1D edge Hamiltonian $H_{\mathrm{1D}}(\delta{k},\theta,\varphi)$.
Before going further, we should make a certain remark simplifying the calculations. With the parameters provided in the Supplementary Materials, one can verify that both $v_x$ and $v_y$ are significantly lower than $v_0$ and $v_z$ for any orientation of the edges (also see Fig.~\ref{Fig:3}). Therefore, one can neglect these terms in the first approximation and take them into account by means of perturbation theory.

Thus, by applying the unitary transformation, the linear part of Eq.~(\ref{eq:9}) can be written as
\begin{equation}
\label{eq:12}
\tilde{H}_{\mathrm{1D}}(\delta{k})
=\varepsilon_0+v_{0}\delta{k}\mathbf{I}_2-v_{z}\delta{k}\sigma_y-m_y\sigma_z-m_x\sigma_x.
\end{equation}
It is clear that Eq.~(\ref{eq:12}) represents a 1D Dirac Hamiltonian with the mass $-m_y$, which changes its sign with $\varphi$ (see Fig.~\ref{Fig:3}), modified by the presence of ``tilted'' term $v_{0}\delta{k}\mathbf{I}_2$ and second mass term $m_x\sigma_x$.

Let us now define the coordinate $x$ along the curved edge so that $x=0$ corresponds to the meeting corner as shown in Fig.~\ref{Fig:3}A. The latter means that $m_x$, $m_y$ in Eq.~(\ref{eq:12}) are the function of $x$, and $\delta{k}=-i\partial/\partial{x}$. Note that $\tilde{H}_{\mathrm{1D}}(-i\partial/\partial{x})$ is defined in disjoint regions out of $x=0$. To fully define the 1D system, one needs to specify the boundary conditions that the wave functions must satisfy in the vicinity of $x=0$.

As shown in the Supplementary Materials, the general \emph{linear} boundary condition, conserving the probability current along the curved edge, can be written in the form
\begin{equation}
\label{eq:13}
e^{-i\beta_1\sigma_y}\Phi_1\big|_{x=-0}=
e^{-i\beta_2\sigma_y}\Phi_2\big|_{x=+0},
\end{equation}
where $\Phi_1$ and $\Phi_2$ are the wave-functions defined from different sides of the corner, while $\beta_1$ and $\beta_2$ are real parameters lying in the range from $-\pi/2$ to $\pi/2$. A physical interpretation of these parameters will be discussed later.

On the basis of Eq.~(\ref{eq:13}), it is convenient to introduce a new Hamiltonian $H_{\mathrm{1D}}^{(\mathrm{new})}=
e^{i\tilde{\beta}\sigma_y}\tilde{H}_{\mathrm{1D}}e^{-i\tilde{\beta}\sigma_y}$, whose eigenfunctions $\Psi_{\mathrm{0D}}(x)$ are continuous at $x=0$:
\begin{multline}
\label{eq:14}
H_{\mathrm{1D}}^{(\mathrm{new})}(\delta{k},x,\tilde{\beta})=\varepsilon_0+v_{0}\hat{k}\mathbf{I}_2-v_{z}\hat{k}\sigma_y+\\
+M_z(x,\tilde{\beta})\sigma_z-M_x(x,\tilde{\beta})\sigma_x,
\end{multline}
where $M_z(x)$ and $M_x(x)$ are
\begin{eqnarray}
\label{eq:15}
M_z(x,\tilde{\beta})=m_x(x)\sin2\tilde{\beta}-m_y(x)\cos2\tilde{\beta},\nonumber\\
M_x(x,\tilde{\beta})=m_x(x)\cos2\tilde{\beta}+m_y(x)\sin2\tilde{\beta}~
\end{eqnarray}
with $\tilde{\beta}$ also being a function of $x$.

In view of the above, the Schr\"{o}dinger equation for the corner states takes the form
\begin{multline}
\label{eq:16}
\left(-v_{z}\delta{k}\sigma_y+M_z\sigma_z-M_x\sigma_x\right)\Psi_{\mathrm{0D}}(x)=\\
=\left(E-\varepsilon_0-v_{0}\delta{k}\right)\mathbf{I}_2\Psi_{\mathrm{0D}}(x).
\end{multline}

An exact solution of Eq.~(\ref{eq:16}) can be found for the case in which $M_z(x)$ and $M_x(x)$ are proportional to each other
\begin{equation}
\label{eq:17}
M_x(x)=\alpha{M_z(x)}+m,
\end{equation}
where $\alpha$ and $m$ are real constants defined as
\begin{eqnarray}
\label{eq:18}
\alpha=\dfrac{M_x(-\infty)-M_x(+\infty)}{M_z(-\infty)-M_z(+\infty)},~~~~~~~~~~~~\nonumber\\
m=\dfrac{M_z(-\infty)M_x(+\infty)-M_z(+\infty)M_x(-\infty)}{M_z(-\infty)-M_z(+\infty)}.
\end{eqnarray}
Note that Eq.~(\ref{eq:17}) is \emph{exact} for the sharp corner shown in Fig.~\ref{Fig:3}, for which $M_x$ and $M_z$ are the step-like functions of $x$.

As shown in the Supplementary Materials, under the condition of Eq.~(\ref{eq:17}), the wave-function can be presented in the form
\begin{equation}
\label{eq:19}
\Psi_{\mathrm{0D}}(x)={\chi}\psi(x),
\end{equation}
where ${\chi}$ is the spin part of the wave function satisfying equation
\begin{equation*}
\begin{pmatrix}
-v_{z}\alpha+iv_{0} & -v_{z}-iv_{0}\alpha \\
-v_{z}-iv_{0}\alpha & v_{z}\alpha-iv_{0}
\end{pmatrix}{\chi}=\nu{\chi},
\end{equation*}
with eigenvalues $\nu=\pm\sqrt{1+\alpha^2}\sqrt{v_{z}^2-v_{0}^2}$.

Then, by introducing a new variable $\tilde{x}=x/\sqrt{v_{z}^2-v_{0}^2}$ and representing $\psi(x)$ in the form
\begin{equation}
\label{eq:20}
\psi(x)=\tilde{\psi}(\tilde{x})e^{\displaystyle{-i\tilde{x}\frac{(E-\varepsilon_0)v_{0}}{\sqrt{v_{z}^2-v_{0}^2}}}},
\end{equation}
we arrive at the equation for the coordinate part:
\begin{equation}
\label{eq:21}
\Bigg\{\hat{\tilde{k}}^2
+\tilde{W}(\tilde{x})^2+\sigma\tilde{W}(\tilde{x})'\Bigg\}\tilde{\psi}(\tilde{x})=\varepsilon\tilde{\psi}(\tilde{x}),
\end{equation}
where $\sigma=\pm{1}$ (the sign of $\sigma$ coincides with those for $\nu$), and $\varepsilon$ and $\tilde{W}(\tilde{x})$ are defined as
\begin{eqnarray}
\label{eq:22}
\varepsilon=\dfrac{(E-\varepsilon_0)^2v_{z}^2}{v_{z}^2-v_{0}^2}-\dfrac{m^2}{1+\alpha^2},~~~\nonumber\\
\tilde{W}(\tilde{x})=\sqrt{1+\alpha^2}M_z+\dfrac{m\alpha}{\sqrt{1+\alpha^2}}.
\end{eqnarray}

Equation~(\ref{eq:21}) possesses a special symmetry that corresponds to the formulation of supersymmetric quantum mechanics~\cite{w52} with the supersymmetric potential $\tilde{W}(\tilde{x})$. If the signs of the asymptotes $\tilde{W}(+\infty)$ and $\tilde{W}(-\infty)$ are opposite, i.e.
\begin{equation}
\label{eq:23}
\left(M_z(+\infty)+\dfrac{m\alpha}{1+\alpha^2}\right)\left(M_z(-\infty)+\dfrac{m\alpha}{1+\alpha^2}\right)<0,
\end{equation}
Eq.~(\ref{eq:21}) always has a localized solution $\tilde{\psi}(\tilde{x})$ with $\varepsilon=0$.

Thus, by means of Eqs~(\ref{eq:20})--(\ref{eq:23}), the wave function of the corner state is written as:
\begin{multline}
\label{eq:24}
\Psi_{\mathrm{0D}}(x)=C
\begin{pmatrix}
v_{z}\alpha-\sigma\sqrt{1+\alpha^2}\sqrt{v_{z}^2-v_{0}^2}-iv_{0} \\
v_{z}+iv_{0}\alpha
\end{pmatrix}\times\\
{\times}e^{\displaystyle\dfrac{\sigma}{\sqrt{1+\alpha^2}\sqrt{v_{z}^2-v_{0}^2}}\int\limits_0^{x}\left\{(1+\alpha^2)M_z(z)+m\alpha\right\}dz}\times\\
{\times}e^{\displaystyle{-ix\dfrac{(E_{\mathrm{0D}}-\varepsilon_0)v_{0}}{v_{z}^2-v_{0}^2}}},
\end{multline}
where $C$ is the normalization constant and
\begin{equation}
\label{eq:25}
E_{\mathrm{0D}}=\varepsilon_0+\dfrac{{\sigma}m}{\sqrt{1+\alpha^2}}
\dfrac{\sqrt{v_{z}^2-v_{0}^2}}{v_z}.
\end{equation}
Note that in Eqs~(\ref{eq:24}) and (\ref{eq:25}), the sign of $\sigma$ should be chosen in accordance with normalized condition of $\Psi_{\mathrm{0D}}(x)$. If $\tilde{W}(+\infty)>0$, $\sigma=-1$, while for $\tilde{W}(+\infty)<0$, $\sigma=1$. These two cases correspond to the internal and external corners at the same orientations of the two edges.

Let us make few remarks concerning the results obtained above. First, Eq.~(\ref{eq:23}) can be also written in equivalent form
\begin{equation}
\label{eq:26}
\left(M_z(+\infty)+{\alpha}M_x(+\infty)\right)\left(M_z(-\infty)+{\alpha}M_x(-\infty)\right)<0,
\end{equation}
which is reduced to the well-known condition for the existence of the bound state in 1D Dirac system if one of the mass parameters $M_z$ or $M_x$ is absent.

Second, $E_{\mathrm{0D}}$ depends on the values of $\tilde{\beta}(-\infty)$ and $\tilde{\beta}(+\infty)$ as seen from Eq.~(\ref{eq:15}). Nevertheless, since
\begin{equation}
\label{eq:27}
(E_{\mathrm{0D}}-\varepsilon_0)^2<m_x(x)^2+m_y(x)^2
\end{equation}
takes place for any values of $\tilde{\beta}(-\infty)$ and $\tilde{\beta}(+\infty)$, the corner state energy always lies in the band-gap of the edge states as soon as Eq.~(\ref{eq:26}) is fulfilled. One can show that $E_{\mathrm{0D}}$ may formally achieve the energies of the 1D band edges at certain values of $\tilde{\beta}^{*}(-\infty)$ and $\tilde{\beta}^{*}(+\infty)$, corresponding to
\begin{equation*}
\left(M_z^{*}(+\infty)+{\alpha^{*}}M_x^{*}(+\infty)\right)\left(M_z^{*}(-\infty)+{\alpha^{*}}M_x^{*}(-\infty)\right)=0.
\end{equation*}
The latter however represents the moment, when the corner state becomes delocalized.

Finally, we take into account the small terms proportional to $v_x$ and $v_y$ previously neglected in Eq.~(\ref{eq:12}). The straightforward calculations on the basis of $\Psi_{\mathrm{0D}}(x)$ (see Supplementary Materials) lead to the first-order energy shift:
\begin{multline}
\label{eq:28}
\delta{E}_{\mathrm{0D}}=-\dfrac{v_{0}m}{v_{z}^2(1+\alpha^2)}\bigg[V(-\infty)\dfrac{\lambda(+\infty)}{\lambda(-\infty)+\lambda(\infty)}+\\
V(+\infty)\dfrac{\lambda(-\infty)}{\lambda(-\infty)+\lambda(\infty)}\bigg],
\end{multline}
where
\begin{eqnarray}
\label{eq:29}
V(x)=(v_x+\alpha{v_y})\cos2\tilde{\beta}+(v_y-\alpha{v_x})\sin2\tilde{\beta},\nonumber\\
\lambda(\pm\infty)=\dfrac{(1+\alpha^2)M_z(\pm\infty)+m\alpha}{\sqrt{1+\alpha^2}\sqrt{v_{z}^2-v_{0}^2}}.~~~~~~~~
\end{eqnarray}
One can verify that indeed $|\delta{E}_{\mathrm{0D}}|\ll|E_{\mathrm{0D}}|$ for both prototype QWs. The calculations for another Kramer's pair of the edge states described by
$H_{\mathrm{1D}}^{*}(-k_x-k_c,\theta,\varphi)$ results in the same energy of the corner state.

\subsection{Trivial corner states}
Let us now discuss the physical origin of the localized 0D corner state found above.
Since it is clear from Eq.~(\ref{eq:15}), the existence condition of Eq.~(\ref{eq:23}) (or equivalently Eq.~(\ref{eq:26})) is fulfilled for different functions $m_x(x)$, $m_y(x)$ and $\tilde{\beta}(x)$. The function $\tilde{\beta}(x)$ characterizes the corner itself, while $m_x$ and $m_y$ include the characteristic of the entire system. Further, we show that, in the most general case, arising of 0D corner state simultaneously depends on the corner boundary conditions and the cubic symmetry of the system.

For a better understanding of the existence condition of Eq.~(\ref{eq:26}), we first neglect the terms arising due to the cubic symmetry. In this case, one should set parameters
$R_{1,2}^{(a)}$ and $\tilde{R}_{1,2}^{(a)}$ of 2D Hamiltonian in Eq.~(\ref{eq:5}) to zero also resulting in the zero values of $F_{a}$ and $\tilde{F}_{a}$ in Eq.~(\ref{eq:11}). The latter, in turn, leads to the vanishing of $m_y$, $v_y$ and $\delta_y$ in Eq.~(\ref{eq:10}), while $m_x$, $v_x$ and $\delta_x$ become independent of $x$. Thus, in the absence of the cubic symmetry, by means of Eqs~(\ref{eq:15}) and (\ref{eq:18}), the existence condition of Eq.~(\ref{eq:26}) is reduced to
\begin{equation}
\label{eq:30}
-\dfrac{4\sin^4(\beta_1-\beta_2)}{\left(\sin2\beta_1-\sin2\beta_2\right)^2}<0,
\end{equation}
where $\beta_1=\tilde{\beta}(x=-\infty)$ and $\beta_2=\tilde{\beta}(x=+\infty)$ chosen in accordance with Eq.~(\ref{eq:13}). Since it is easy to see, the localized corner state exists if $\beta_1\neq\beta_2$. These parameters can be given a precise physical interpretation.

Let us consider a $\delta$-function electrostatic potential at $x=0$. Then, if we integrate the Schr\"{o}dinger equation with $\tilde{H}_{\mathrm{1D}}(\delta{k},x)+V_0\delta(0)\mathbf{I}_2$ (where $V_0$ is a real parameter) in the range of $-\eta{\leq}x{\leq}\eta$ (where $\eta$ is a positive infinitesimal quantity), we directly obtain Eq.~(\ref{eq:13}) with $\beta_2-\beta_1=V_0v_z/(v_z^2-v_0^2)$. Therefore, $\beta_1\neq\beta_2$ can be interpreted as the presence of $\delta$-like electrostatic potential localized at the corner. Thus, the corner states arising in this case are topologically trivial, since they are independent of the cubic symmetry of the system. We also note their independence of the sign of $V_0$, as clearly seen from Eq.~(\ref{eq:30}).

\subsection{Topological corner states}
We now focus on the opposite case, when there is no potential barrier at the corner i.e. $\beta_1=\beta_2$. As seen from Eq.~(\ref{eq:13}), these parameters can be both set to zero without loss of generality, which leads to $M_x(x,0)=m_x(x)$ and $M_z(x,0)=-m_y(x)$. In this case, the presence of the corner states is governed by the cubic symmetry of 2D system, represented by the non-zero values of $F_{a}$ and $\tilde{F}_{a}$ in Eq.~(\ref{eq:11}). The latter means that the picture of the symmetry-protected corner states should strongly depends not only on crystallographic orientations of the meeting edges but also on the QW growth direction.

\begin{figure}
\includegraphics [width=1.0\columnwidth, keepaspectratio] {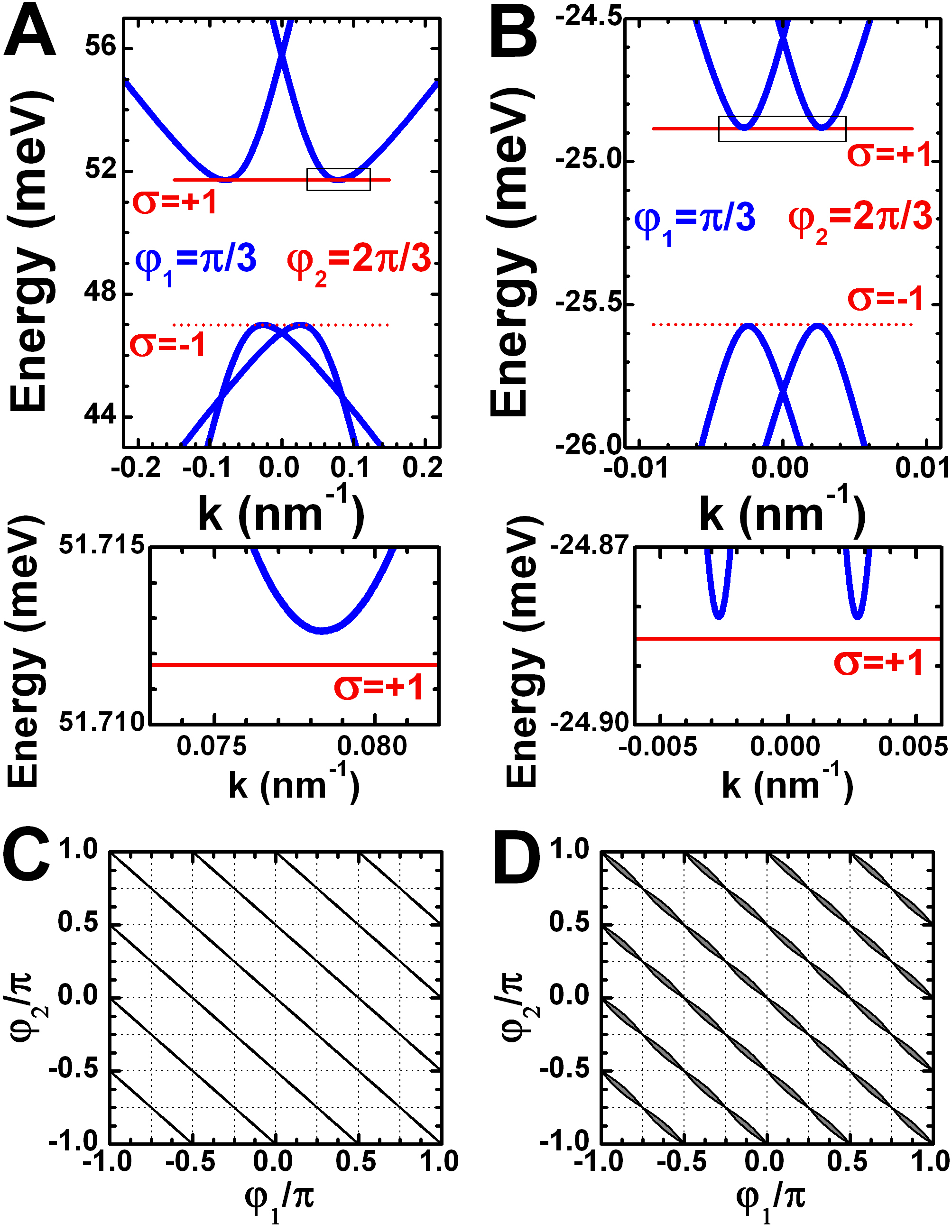} 
\caption{\label{Fig:4} \textbf{Topological corner states in (001)-oriented prototype QWs.}
(\textbf{A,B}) 1D dispersion of the edge states (in blue) for the boundary oriented at $\varphi_1=\pi/3$ and the corner states energy $E_{\mathrm{0D}}+{\delta}E_{\mathrm{0D}}$ (in red) formed by the second boundary oriented at $\varphi_2=2\pi/3$. The A and B panels correspond to the (001) three-layer InAs/GaInSb (A) and double HgTe/CdHgTe QWs, respectively. The layer thicknesses of the QWs are marked by the red symbols in Figs~\ref{Fig:1}B and \ref{Fig:1}C.
(\textbf{C,D}) The phase diagram for the presence (grey region) and absence (white region) of topological corner states in the three-layer InAs/GaInSb and double HgTe/CdHgTe QWs as a function of the two edge orientations. In the grey regions, the existence condition of Eq.~(\ref{eq:26}) at $\beta_1=\beta_2=0$ is fulfilled. The vertical and horizontal thin dotted lines represent the angles when the edges of 2D system, which are actually the faces of (001)-oriented QWs, coincide with $[110]$, $[1\bar{1}0]$, $[100]$ and $[010]$ crystallographic planes.}
\end{figure}

Figures~\ref{Fig:4}A and \ref{Fig:4}B present the calculations for the prototype QWs with double band inversion grown along (001) crystallographic orientation (see the diagrams in Fig.~\ref{Fig:1}). For both QWs, the angles $\varphi_1$ and $\varphi_2$ defining the corner orientation in Fig.~\ref{Fig:3}A are chosen to be $\varphi_1=\pi/3$ and $\varphi_2=2\pi/3$, while $\theta$ in Eq.~(\ref{eq:10}) is assumed to be zero. One can verify that the condition of existence of Eq.~(\ref{eq:26}) is fulfilled for these angles, and, therefore, the corner states arise in the system. As seen from Figs~\ref{Fig:4}A and \ref{Fig:4}B, due to the relationship between $m_x(\varphi)$ and $m_y(\varphi)$, the localized energies for internal $\sigma=+1$ and external $\sigma=-1$ corners are very close to the extrema of 1D dispersion of the edge states.

Figures~\ref{Fig:4}C and \ref{Fig:4}D represent the angle diagram showing the edge orientations in the (001)-oriented prototype QWs, for which the existence condition of Eq.~(\ref{eq:26}) is valid. It is seen that the values of $\varphi_1$ and $\varphi_2$ yielding the corner states are qualitatively represented as the grey regions elongated along the lines defined by
\begin{equation}
\label{eq:31}
\varphi_1+\varphi_2=n_0\left(2\pi/4\right),
\end{equation}
excluding the points lying at verticals and horizontals corresponding to $\varphi_1=m_0\pi/4$ and $\varphi_2=l_0\pi/4$, where $n_0$, $m_0$ and $l_0$ are integers. As seen from Fig.~\ref{Fig:3}A, the 2D system edges at these angles, coincides with $[110]$, $[1\bar{1}0]$, $[100]$ or $[010]$ crystallographic planes, being the faces of (001)-oriented QWs in this case. Since we have neglected the terms resulting from possible breaking of inversion symmetry in the system, the (001)-oriented prototype QWs preserve the mirror symmetry about the mentioned planes. Therefore, one concludes that if one of the QW edge coincides with the mirror symmetry planes, the corner states are absent in the system.

\begin{figure}
\includegraphics [width=1.0\columnwidth, keepaspectratio] {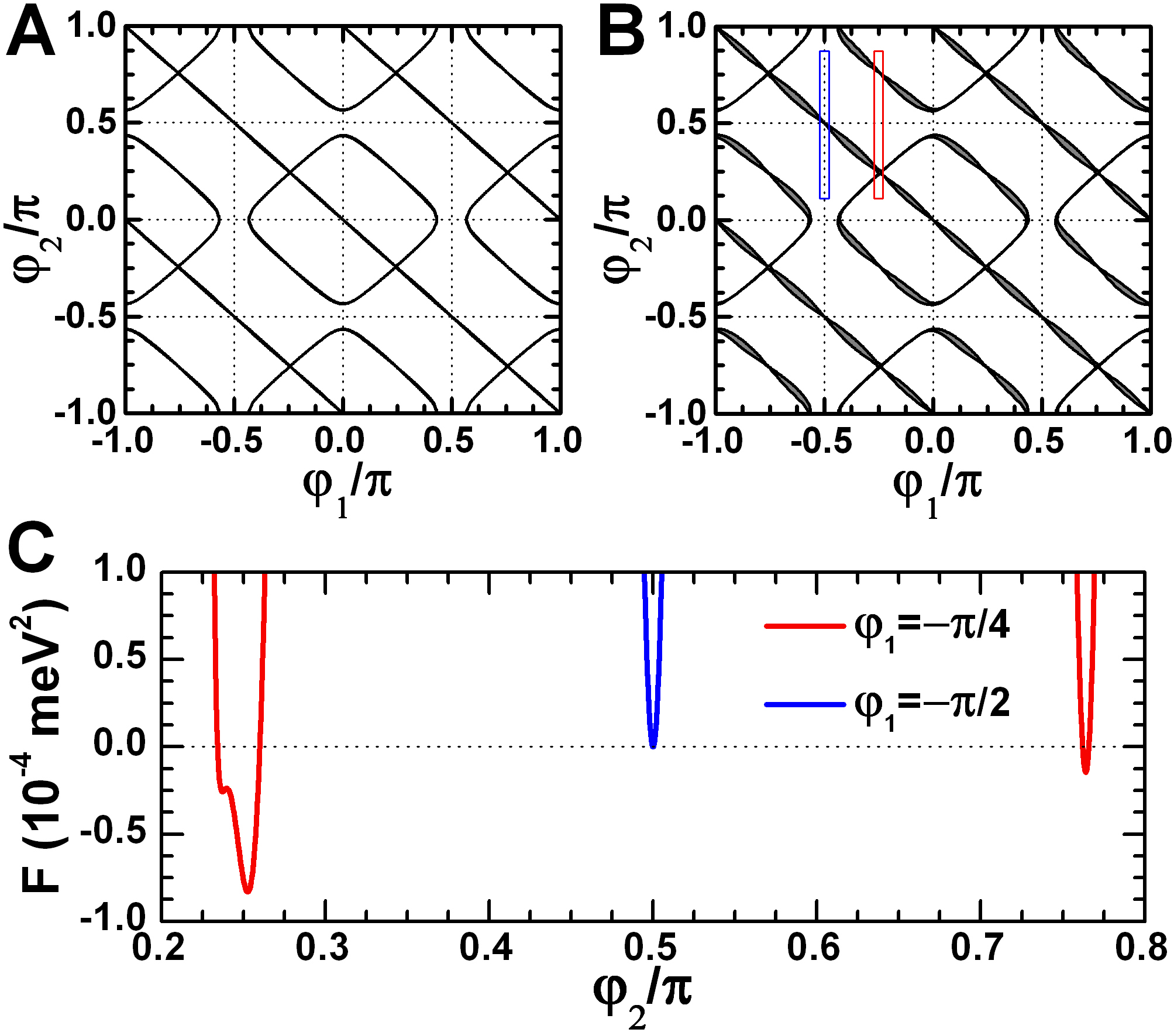} 
\caption{\label{Fig:5} \textbf{Topological corner states in ($013$)-oriented prototype QWs.}
(\textbf{A,B}) The phase diagram for the presence (grey region) and absence (white region) of 0D topological corner states in the three-layer InAs/GaInSb (A) and double HgTe/CdHgTe QWs (B) as a function of the two edge orientations. The vertical and horizontal thin dotted lines represent the angles when the edges of 2D system coincide with $[100]$ and $[03\bar{1}]$ crystallographic planes. (\textbf{C}) The function $F(\varphi_1,\varphi_2)=\left(M_z(\varphi_1)+{\alpha}M_x(\varphi_1)\right)\left(M_z(\varphi_2)+{\alpha}M_x(\varphi_2)\right)$
representing the existence condition of Eq.~(\ref{eq:26}) at $\beta_1=\beta_2=0$ in the ranges marked with colored rectangles in the panel (B). The dotted line marks the zero value.}
\end{figure}

In light of the above, Eq.~(\ref{eq:31}) describes the corners, whose bisector corresponds to the one of the mirror symmetry planes. The latter results in reflection symmetry of the angle diagram with respect to the lines described by Eq.~(\ref{eq:31}). As expected from the four-fold rotational symmetry of (001)-oriented QWs with respect to the growth direction, the angle pattern in Figs~\ref{Fig:4}C and \ref{Fig:4}D also possess a $\pi/2$-periodicity in $\varphi_1$ and $\varphi_2$.

So far, we have discussed the case of (001)-oriented QWs, by whose example we have shown a direct relationship between topological corner states and the symmetry elements of the QW. As the inversion symmetry is preserved in our model, the (001)-oriented QWs have the point group $D_{4h}$ origin from the point group $O_h$ of bulk multi-band \textbf{k$\cdot$p} Hamiltonian (see Supplementary Materials). Obviously, if one reduces the symmetry of 2D system, the picture of the corner states should change as well. Further, we consider $C_{2h}$-symmetric QWs grown along ($013$) crystallographic direction inspired by recent experimental investigations of double HgTe QWs of the same orientation~\cite{w53,w54,w55}.

Figures~\ref{Fig:5}A and \ref{Fig:5}B provide the angle diagram of topological corner states in the prototypes three-layer InAs/GaInSb and double HgTe/CdHgTe QWs oriented along ($013$) crystallographic direction. The calculations are performed by assuming $\theta=\arctan(m/n)$ in Eq.~(\ref{eq:10}) with ($m$,~$n$)=($1$,~$3$). It is seen that the values of $\varphi_1$ and $\varphi_2$, for which the existence condition is fulfilled, represent significantly different picture of the corner states than the one for (001)-oriented QWs.

We first note $\pi$-periodicity in $\varphi_1$ and $\varphi_2$ of the angle pattern due to the two-fold rotational symmetry of ($0mn$)-oriented QWs in contrast to the $\pi/2$-periodicity of (001)-oriented 2D systems. Second, due to the symmetry lowering, (013)-oriented QWs have only two $[100]$ and $[03\bar{1}]$ mirror planes perpendicular to the QW plane. These planes are represented by dotted verticals and horizontals at $\varphi_1=m_0\pi/2$ and $\varphi_2=l_0\pi/2$ in Figs~\ref{Fig:5}A and \ref{Fig:5}B. It is seen that the corner states are absent if one of the QW edge coincides with the mirror symmetry planes. Finally, in contrast to (001)-oriented QWs, ($0mn$)-oriented QWs possess the corner states if one of the edges is at $\pi/4$ degrees with respect to the ($100$) crystallographic axis. The latter is clearly seen in Fig.~\ref{Fig:5}C, which shows the negative values of the function illustrating the existence condition of Eq.~(\ref{eq:26}) for the corner states.

\section{Discussion}
To summarize, we have investigated the existence conditions for 0D corner states in cubic semiconductor QWs with double band inversion. We have demonstrated that 0D corner states in such 2D system can appear either due to the presence of electrostatic potential localized in the corner, or due to the crystalline symmetry of the QW. The former case can be physically interpreted as the impurity potential resulting in trivial corner states. The latter corresponds to the symmetry-induced corner states inherent for 2D HOTI state in cubic semiconductor QWs with double band inversion. We have shown that the corners hosting 0D topological states depend not only on crystallographic orientations of the meeting edges but also on the growth orientation and parameters of the QWs.

Let us now make a few remarks concerning the approximation used in this work. As it is clear from the above, theoretical investigations were performed on the basis of multi-band \textbf{k$\cdot$p} Hamiltonian~\cite{w46} for the envelope Bloch functions, which is indeed valid on scales much larger than the unit cell. In this case, all functions that are changed on a scale comparable to the unit cell are considered as step-like or delta-functions. In this sense, a meeting point of the corner shown in Fig.~\ref{Fig:3}A represents a 2D unit cell at the corner of an effective 2D system in the QW plane. The special arrangement of atoms in the vicinity of the corner preserving $C_n$ symmetry of the bulk may lead to the fractional charge of the corner states, known as a fractional corner anomaly~\cite{w15b,w15c,w58,w59,w60}. We note that since fractional corner anomaly requires a theoretical description on the unit cell scale, it cannot be treated on the basis of multi-band \textbf{k$\cdot$p} Hamiltonian.

Throughout the work, we have neglected the inversion symmetry breaking terms resulting from anisotropy of chemical bonds at the QW interfaces~\cite{w49} and possible bulk inversion asymmetry of the unit cell~\cite{w50}. The former is known as an interface inversion asymmetry~\cite{w49}, while the latter causes the difference between $O_h$ and $T_d$ point groups of diamond and zinc-blende semiconductors. Taking into account these additional terms in bulk \textbf{k$\cdot$p} Hamiltonian leads to the non-diagonal blocks of the effective 2D Hamiltonian $H_{2D}(k_x,k_y)$ in Eq.~(\ref{eq:1}). Since the inversion symmetry breaking terms do not affect the very fact of double band inversion, their presence will modify only the edge state parameters $v_x$, $v_y$, $m_x$, $m_y$, $\delta_x$, $\delta_y$ keeping the form of 1D edge Hamiltonian in Eq.~(\ref{eq:9}).

For instance, taking into account the interface and bulk inversion asymmetry in (001)-oriented zinc-blende semiconductor QW reduces the point symmetry from $D_{4h}$ to $C_{2v}$~\cite{w49}, which, in their turn, also changes the periodicity of $v_x(\varphi)$, $v_y(\varphi)$, $m_x(\varphi)$, $m_y(\varphi)$, $\delta_x(\varphi)$, $\delta_y(\varphi)$ in Eq.~(\ref{eq:9}) from $\pi/2$ to $\pi$. In this case, the symmetry-induced corner states will arise at the corner with the meeting edges, whose orientations differ from those shown in Fig.~\ref{Fig:4}. For the prototype InAs/GaInSb and HgTe/CdHgTe QWs considered in this work, the terms due to interface and bulk inversion asymmetry are small and induce therefore only a slight modification of the phase diagrams.

We emphasize that since the presence of the corner states depends on the mutual ratio between $m_x$ and $m_y$, one may find the QW orientation and strength of inversion-symmetry breaking terms, for which the existence condition of Eq.~(\ref{eq:26}) cannot be fulfilled.
The study of all possible cases of violation of Eq.~(\ref{eq:26}) is however out the scope of our work first considering the QWs based on IV, II-VI and III-V semiconductors, in which inversion-symmetry breaking terms are usually small.

Finally, we stress the importance of theoretical results obtained in this work in view of possible applications and their impact on further experimental investigations. After the tremendous interest in II-VI and III-V semiconductor QWs induced by prediction and observation of QSHI~\cite{w34,w35,w36}, our work shows the importance of cubic semiconductor QWs for the realization of high-order topological states as well.
In view of mature growth of IV, II-IV and III-V semiconductor QWs on Si-wafers~\cite{w61,w62} as well as device fabrication technology, our results provide an important first step to future realistic electronics operating on the basis of higher-order topological states including higher-order topological superconductors in hybrid devices~\cite{w56,w57}.

\section{METHODS}
Band structure calculations were performed by using multi-band \textbf{k$\cdot$p} Hamiltonian~\cite{w46}, which directly takes into account the interactions between $\Gamma_6$, $\Gamma_8$, and $\Gamma_7$ bands in bulk materials. This model well describes the electronic states in a wide range of narrow-gap semiconductor QWs, particularly in the InAs/GaInSb~\cite{w41,w43} and HgTe/CdHgTe QWs~\cite{w53,w54,w55}. In the multi-band \textbf{k$\cdot$p} Hamiltonian, we also took into account the terms, describing the strain effect arising because of the mismatch
of lattice constants in the buffer, QW layers, and barriers. The calculations had been performed
by expanding the eight-component envelope wave functions in the basis set of plane waves and by numerical solution of the eigenvalue problem. Details of calculations and the form of the Hamiltonian can be found in the study of Krishtopenko~et~al.~\cite{w46}. Parameters for the bulk materials and valence band offsets for the the InAs/GaInSb and HgTe/CdHgTe QWs used in the calculations are provided in Ref.~\cite{w45} and Ref.~\cite{w46}, respectively.

To derive effective 2D Hamiltonian valid in the vicinity of the $\Gamma$ point from the multi-band \textbf{k$\cdot$p} Hamiltonian, we implied the procedure proposed by Bernevig et al.~\cite{w2} and described in details in the Supplementary Materials.

~\\~

\begin{acknowledgments}
The author is grateful to Fr\'{e}d\'{e}ric Teppe from Laboratoire Charles Coulomb for helpful discussions on the manuscript. This work was partially supported by the Foundation for Polish Science: the IRAP program (Grant No. MAB/2018/9, project CENTERA), by MIPS department of Montpellier University through the ``Occitanie Terahertz Platform'', by CNRS through IRP ``TeraMIR'' and by the French Agence Nationale pour la Recherche (Colector project).
\end{acknowledgments}


%

\newpage
\clearpage
\setcounter{equation}{0}
\setcounter{figure}{0}
\setcounter{table}{0}

\onecolumngrid
\begin{center}
\LARGE{\textbf{Supplementary Materials}}
\end{center}
\maketitle
\onecolumngrid

\renewcommand{\thefigure}{S\arabic{figure}} 
\renewcommand{\thetable}{S\arabic{table}}   

This PDF file includes:
\begin{itemize}
\item Supplementary Text;
\item Fig.~\ref{Fig:SM1}. Parameters of effective 1D edge Hamiltonian as a function of the edge orientation $\varphi$ for the three-layer InAs/GaInSb and double HgTe/CdHgTe QWs considered in the main text at different growth orientations. The edge independent parameters ($v_0$, $v_z$) equal to ($36.6$, $68.1$)~meV$\cdot$nm and ($-16.9$, $200.3$)~meV$\cdot$nm for the three-layer InAs/GaInSb and double HgTe/CdHgTe QWs, respectively;
\item Table~\ref{tab:1}. Parities of the envelope functions of multi-band \textbf{k$\cdot$p} Hamiltonian at zero electron momentum in the QW plane;
\item Table~\ref{tab:2}. Parameters involved in the effective 2D Hamiltonian for the three-layer InAs/GaInSb and double HgTe/CdHgTe QWs considered in the main text.
\end{itemize}

\newpage
\section*{Supplementary Text}
\subsection{An effective four-band model for cubic-semiconductor quantum well grown along [0$mn$] direction}
A correct theoretical description of 2D systems with double band inversion is possible only on the basis of a Hamiltonian, which takes into account at least four bands. The minimal required effective 2D four-band Hamiltonian was first proposed for (001) double HgTe/CdHgTe QWs~\cite{HOTIsm9}. It was derived from the eight-band Kane model for the envelope wave functions of the $\Gamma_6$, $\Gamma_8$ and $\Gamma_7$ bands by preserving the full rotational symmetry in the QW plane~\cite{HOTIsm11,HOTIsm12} and omitting the warping terms. In this section, we go beyond the axial approximation of Refs~\cite{HOTIsm9,HOTIsm11,HOTIsm12} and include the warping terms origin from cubic symmetry of diamond and zinc-blende semiconductors. The latter is essential for higher-order topology of cubic semiconductor QWs with double band inversion. In order to demonstrate that occurrence of the corner states depends on the growth orientation of 2D system, we further consider the general class of ($0mn$)-oriented QWs~\cite{HOTIsm10,HOTIsm10b} ($m$ and $n$ are integer numbers), which includes (001)-, (011)-, and (013)-oriented structures.

For simplicity, we limit ourselves to the upper $6\times6$ block of the Kane model including only the $\Gamma_6$ and $\Gamma_8$ bands. Although the $\Gamma_7$ band is needed for quantitative description of the positions of electron-like subbands in narrow QWs (see Supplemental material of Ref.~\cite{HOTIsm10}), it can be ignored for qualitative description of the band structure. We note that taking into account the contribution of the $\Gamma_7$ band does not lead to any new terms and just complicates the derivation of the effective 2D Hamiltonian (cf.~Supplemental materials of Ref.~\cite{HOTIsm9} and Ref.~\cite{HOTIsm13}). Further, we also neglect the terms breaking inversion symmetry of 2D systems. These terms result from (i) the absence of inversion center in the unit cell of zinc-blende semiconductors; (ii) from the anisotropy of chemical bonds at the QW interfaces leading to the interface inversion asymmetry; and (iii) from the structural inversion asymmetry (SIA) in the growth direction. The latter arises only if the QW profile is asymmetric.

In the basis set of Bloch amplitudes in the sequence $|\Gamma_6,+1/2\rangle$,$|\Gamma_6,-1/2\rangle$, $|\Gamma_8,+3/2\rangle$, $|\Gamma_8,+1/2\rangle$, $|\Gamma_8,-1/2\rangle$, $|\Gamma_8,-3/2\rangle$~\cite{HOTIsm10}, the 6-band \textbf{k$\cdot$p} Hamiltonian preserving inversion asymmetry is given by
\begin{equation}
\label{eq:1DH1}
H_{3D}=\begin{pmatrix}
H_{cc} & H_{cv} \\H_{cv}^{\dag} & H_{vv}\end{pmatrix},
\end{equation}
where the blocks $H_{cc}$ and $H_{vv}$ represent the contribution from the $\Gamma_6$ and $\Gamma_8$ bands, respectively, and the block $H_{cv}$ and the block $H_{cv}$ describes the band mixing. The block $H_{cc}$ is given by
\begin{equation}
\label{eq:1DH2}
H_{cc}=\left[E_c(z)+\dfrac{\hbar^{2}\mathbf{k}\left[2F(z)+1\right]\mathbf{k}}{2m_0} + \Xi_{c}\mathrm{Tr}\epsilon\right]I_{2{\times}2},
\end{equation}
where $I_{2{\times}2}$ is the $2{\times}2$ identity matrix, $E_c(z)$ is the conduction band profile, $\mathbf{k}=(k_x,k_y,k_z)$ (note that $k_z=-i\partial/\partial{z}$ as $z$ is the growth direction), $F(z)$ is a parameter accounting for contribution from remote bands, $\Xi_{c}$ is the $\Gamma_6$-band deformation potential constant, and $\epsilon$ is the train tensor arising due to lattice-mismatch in the QW layers and the sample substrate. The block $H_{cv}$ has the form
\begin{equation}
\label{eq:1DH3}
H_{cv}=\begin{pmatrix}
-\dfrac{\sqrt{2}Pk_{+}}{2} & \dfrac{\sqrt{6}Pk_{z}}{3} & \dfrac{\sqrt{6}Pk_{-}}{6} & 0 \\[6pt]
0 & -\dfrac{\sqrt{6}Pk_{+}}{6} & \dfrac{\sqrt{6}Pk_{z}}{3} & \dfrac{\sqrt{2}Pk_{-}}{2} \end{pmatrix},
\end{equation}
where $P$ is the Kane matrix element, $k_{\pm}=k_x{\pm}ik_y$. The block $H_{vv}$ is given by
\begin{equation}
\label{eq:1DH5}
H_{vv}=E_v(z)I_{4{\times}4}+H^{(i)}_L+H^{(a)}_L+H^{(i)}_{BP}+H^{(a)}_{BP},
\end{equation}
where $I_{4{\times}4}$ is the $4{\times}4$ identity matrix, $E_v(z)$ is the valence band profile, $H^{(i)}_L$, $H^{(a)}_L$, $H^{(i)}_{BP}$ and $H^{(a)}_{BP}$ are the isotropic and anisotropic parts of the Luttinger and Bir-Pikus Hamiltonians,
\begin{eqnarray}
\label{eq:1DH6}
H^{(i)}_L=\dfrac{\hbar^2}{2m_0}\left[-\mathbf{k}\left(\gamma_1+\dfrac{5}{2}\gamma_2\right)\mathbf{k}+ 2(\mathbf{J}\cdot\mathbf{k})\gamma_2(\mathbf{J}\cdot\mathbf{k})\right],\nonumber\\
H^{(i)}_{BP}=\left(a+\dfrac{5}{4}b\right)\mathrm{Tr}\epsilon-b\sum_{\alpha}J_\alpha^2\epsilon_{\alpha\alpha} -b\sum_{\alpha\neq\beta}\{J_\alpha,J_\beta\}_s\epsilon_{\alpha\beta},
\end{eqnarray}
$\gamma_1$, $\gamma_2$, and $\gamma_3$ are contributions to the Luttinger parameters from remote bands; $\textbf{J}$ is the vector composed of the matrices of the angular momentum $3/2$; $\{J_\alpha,J_\beta\}_s=(J_{\alpha}J_{\beta}+J_{\beta}J_{\alpha})/2$; $a$, $b$, and $d$ are the $\Gamma_8$-band deformation potential constants.

In the coordinate frame relevant to ($0mn$)-oriented QWs $x\parallel[100]$, $y\parallel[0n\bar{m}]$, and $z\parallel[0mn]$ the terms $H^{(a)}_L$ and $H^{(a)}_{BP}$ assume the form~\cite{HOTIsm10,HOTIsm10b}:
\begin{multline*}
H^{(a)}_L=\dfrac{\hbar^2}{2m_0}\bigg(\{J_x,J_y\}_s(\gamma_3-\gamma_2)k_{x}k_{y} +\{J_x,J_z\}_s\{\gamma_3-\gamma_2,k_{z}\}_{s}k_{x} + \\
\left[\{J_y,J_z\}_s\cos{2\theta}-\dfrac{J_z^2-J_y^2}{2}\sin{2\theta}\right]
\left[\{\gamma_3-\gamma_2,k_{z}\}_sk_{y}\cos{2\theta}-\dfrac{k_{z}(\gamma_3-\gamma_2)k_{z}-(\gamma_3-\gamma_2)k_{y}^2}{2}\sin{2\theta}\right] \bigg),~~~~~~~~
\end{multline*}
\begin{multline}
\label{eq:1DH7}
H^{(a)}_{BP}=-2\left(\dfrac{d}{\sqrt{3}}-b\right)\bigg(\{J_x,J_y\}_s\epsilon_{xy} +\{J_x,J_z\}_s\epsilon_{xz} + \\
\left[\{J_y,J_z\}_s\cos{2\theta}-\dfrac{J_z^2-J_y^2}{2}\sin{2\theta}\right]
\left[\epsilon_{yz}\cos{2\theta}-\dfrac{\epsilon_{zz}-\epsilon_{yy}}{2}\sin{2\theta}\right] \bigg),~~~~
\end{multline}
where $\theta=\arctan(m/n)$ is angle between the growth direction [$0mn$] and the [$001$] axis. Expressions for the strain tensor components $\epsilon_{\alpha\beta}$ for the ($0mn$)-oriented QWs are provided in the Supplemental materials of Ref.~\cite{HOTIsm10}.

We now derive an effective four-band 2D Hamiltonian for zinc-blende QW. Following the procedure described in Refs.~\cite{HOTIsm13,HOTIsm14}, we split $H_{3D}$ in Eq.~(\ref{eq:1DH1}) into two parts
\begin{equation}
\label{eq:1DH10}
H_{3D}=H_0(k_z)+H_1(k_z,k_x,k_y,\theta),
\end{equation}
where $H_0(k_z)$ preserves the inversion symmetry \textbf{P} at $k_x=k_y=0$ and $\theta=0$ and the full rotational symmetry in the QW plane, while $H_1(k_z,k_x,k_y,\theta)$ includes the rest terms.

First, we diagonalize the Hamiltonian $H_0(k_z)$ to obtain the energies and envelope functions of the QW states, as well as to classify of electronic levels as electron-like \emph{E}$n$, hole-like \emph{H}$n$ or light-hole-like \emph{LH}$n$ levels ($n=1,2\ldots$). Since it is clear from the form of $H_0(k_z)$, the hole-like levels \emph{H}$n$ at $k_x=k_y=0$ are decoupled from the \emph{E}$n$ and \emph{LH}$n$. Therefore, the eigenfunctions of $H_0(k_z)$ are expanded in the basis of Bloch amplitudes as
\begin{eqnarray}
\label{eq:1DH11}
|En,+\rangle=\begin{pmatrix}
f_1^{(En)}(z)|\Gamma_6,+1/2\rangle \\
0 \\
0 \\
f_4^{(En)}(z)|\Gamma_8,+1/2\rangle \\
0 \\
0
\end{pmatrix},~~
|En,-\rangle=\begin{pmatrix}
0 \\
f_2^{(En)}(z)|\Gamma_6,-1/2\rangle \\
0 \\
0 \\
f_5^{(En)}(z)|\Gamma_8,-1/2\rangle \\
0
\end{pmatrix},~~
|LHn,+\rangle=\begin{pmatrix}
f_1^{(LHn)}(z)|\Gamma_6,+1/2\rangle \\
0 \\
0 \\
f_4^{(LHn)}(z)|\Gamma_8,+1/2\rangle \\
0 \\
0
\end{pmatrix},~~~~\nonumber\\
|LHn,-\rangle=\begin{pmatrix}
0 \\
f_2^{(LHn)}(z)|\Gamma_6,-1/2\rangle \\
0 \\
0 \\
f_5^{(LHn)}(z)|\Gamma_8,-1/2\rangle \\
0
\end{pmatrix},~~
|Hn,+\rangle=\begin{pmatrix}
0 \\
0 \\
f_3^{(Hn)}(z)|\Gamma_8,+3/2\rangle \\
0 \\
0 \\
0
\end{pmatrix},~~
|Hn,-\rangle=\begin{pmatrix}
0 \\
0 \\
0 \\
f_4^{(LHn)}(z)|\Gamma_8,-3/2\rangle \\
0 \\
0
\end{pmatrix}.~~~~
\end{eqnarray}

The presence of time reversal symmetry $\mathcal{T}$ and inversion symmetry \textbf{P} in $H_0(k_z)$ yields the relations~\cite{HOTIsm14}:
\begin{eqnarray*}
f_2^{(E\{2k+1\})}=\left(f_1^{(E\{2k+1\})}\right)^{*},~~~~~ f_5^{(E\{2k+1\})}=-\left(f_4^{(E\{2k+1\})}\right)^{*}, \\
f_2^{(E\{2k+2\})}=-\left(f_1^{(E\{2k+2\})}\right)^{*},~~~~~ f_5^{(E\{2k+2\})}=\left(f_4^{(E\{2k+2\})}\right)^{*}, \\
f_6^{(H\{2k+1\})}=\left(f_3^{(H\{2k+1\})}\right)^{*}, ~~~~~~~~~~~~~~~~~~~~~~\\ f_6^{(H\{2k+2\})}=-\left(f_3^{(H\{2k+2\})}\right)^{*}, ~~~~~~~~~~~~~~~~~~~\\
f_2^{(LH\{2k+1\})}=-\left(f_1^{(LH\{2k+1\})}\right)^{*},~~~~~ f_5^{(LH\{2k+1\})}=\left(f_4^{(LH\{2k+1\})}\right)^{*}, \\
f_2^{(LH\{2k+2\})}=\left(f_1^{(LH\{2k+2\})}\right)^{*},~~~~~ f_5^{(LH\{2k+2\})}=-\left(f_4^{(LH\{2k+2\})}\right)^{*}.
\end{eqnarray*}
Time reversal symmetry relates states with opposite spin; hence when the effective Hamiltonian for one spin is constructed, the Hamiltonian for the opposite spin can be easily obtained through the operation $\mathcal{T}$. The inversion operation \textbf{P} defines the parity of each subband. Since $H_0(k_z)$ preserves inversion symmetry, their eigenstates are also the eigenstates of the inversion operation \textbf{P}, which can greatly simplify the calculation of the matrix elements of the effective 2D Hamiltonian. The parity of each subband is determined by both the envelope functions and the Bloch amplitudes at the $\Gamma$ point. The parities of the envelope functions obtained by diagonalization of $H_0(k_z)$~\cite{HOTIsm13,HOTIsm14} are summarized in Table~\ref{tab:1}.
The parities of the Bloch amplitudes are given by $\mathbf{P}|\Gamma_6,{\pm}1/2\rangle=-|\Gamma_6,{\pm}1/2\rangle$, $\mathbf{P}|\Gamma_8,{\pm}1/2\rangle=|\Gamma_8,{\pm}1/2\rangle$ and $\mathbf{P}|\Gamma_8,{\pm}3/2\rangle=|\Gamma_8,{\pm}3/2\rangle$. Thus, the parities of the QW subbands are $\mathbf{P}|E\{2k+1\},{\pm}\rangle=-|E\{2k+1\},{\pm}\rangle$, $\mathbf{P}|E\{2k+2\},{\pm}\rangle=|E\{2k+2\},{\pm}\rangle$, $\mathbf{P}|H\{2k+2\},{\pm}\rangle=|H\{2k+2\},{\pm}\rangle$, $\mathbf{P}|H\{2k+2\},{\pm}\rangle=-|H\{2k+2\},{\pm}\rangle$, $\mathbf{P}|LH\{2k+2\},{\pm}\rangle=|LH\{2k+2\},{\pm}\rangle$ and $\mathbf{P}|LH\{2k+2\},{\pm}\rangle=-|LH\{2k+2\},{\pm}\rangle$.

Next, we group the eigenstates of Eq.~(\ref{eq:1DH11}) into two classes. The first class, marked as class A, includes the basis states of our final effective 2D Hamiltonian $|E1{\pm}\rangle$, $|H1{\pm}\rangle$, $|H2{\pm}\rangle$ and $|E2{\pm}\rangle$. In the second class, denoted as class~B, we consider all other subbands of the QW. The states in both classes are not coupled, since they are eigenstates of Hamiltonian $H_0(k_z)$. However, the presence of $H_1(k_z,k_x,k_y,\theta)$ introduces the mixing between the states from classes A and B. To derive the effective 2D Hamiltonian, we treat $H_1(k_z,k_x,k_y,\theta)$ as a small perturbation and perform a unitary transformation~\cite{HOTIsm6} to eliminate the coupling between the states from different classes by applying the second-order perturbation formula
\begin{equation}
\label{eq:1DH12}
H_{2D}(k_x,k_y)_{m,m'}=E_{m}\delta_{m,m'}+H'_{m,m'}+
\dfrac{1}{2}\sum\limits_{l}H'_{m,l}H'_{l,m'} \left(\dfrac{1}{E_{m}-E_{l}}+\dfrac{1}{E_{m'}-E_{l}}\right),
\end{equation}
where
\begin{eqnarray}
\label{eq:1DH13}
E_{m}=\int\limits_{-\infty}^{+\infty}dz\sum\limits_{\alpha,\beta=1}^{6} f_\alpha^{(m)}(z)^{*}\left(H_0(k_z)\right)_{\alpha,\beta}f_\beta^{(m')}(z),\nonumber\\
H'_{m,m'}=\int\limits_{-\infty}^{+\infty}dz\sum\limits_{\alpha,\beta=1}^{6} f_\alpha^{(m)}(z)^{*}\left(H_1(k_z,k_x,k_y,\theta)\right)_{\alpha,\beta}f_\beta^{(m')}(z).
\end{eqnarray}
Here, the summation indices $m$, $m'$ correspond to the states in class~A, while the index $l$ is for the states in class~B. The Greek indices label envelope function component of $H_0(k_z)$. We note that accounting for the parity of the envelope functions $f_\alpha^{(m)}(z)$ given in Table~\ref{tab:1} greatly simplifies calculation of $H'_{m,m'}$~\cite{HOTIsm14}.

Ordering the basis states as $|E1{+}\rangle$, $|H1{+}\rangle$, $|H2{-}\rangle$, $|E2{-}\rangle$, $|E1{-}\rangle$, $|H1{-}\rangle$, $|H2{+}\rangle$, $|E2{+}\rangle$, after calculating the matrix-elements in Eqs.~(\ref{eq:1DH12}) and (\ref{eq:1DH13}), we are left with the effective 2D Hamiltonian in the form:
\begin{equation}
\label{eq:1DH14}
H_{2D}(k_x,k_y,\theta)=\begin{pmatrix}
H_{4\times4}(k_x,k_y,\theta) & 0 \\ 0 & H_{4\times4}^{*}(-k_x,-k_y,\theta)\end{pmatrix}.
\end{equation}
The diagonal blocks $H_{4\times4}(k_x,k_y,\theta)$ and $H_{4\times4}^{*}(-k_x,-k_y,\theta)$, in their turn, $H_{4\times4}(\mathbf{k},\theta)$ can be split into isotropic and anisotropic parts:
\begin{equation}
\label{eq:1DH15}
H_{4\times4}(k_x,k_y,\theta)=H_{4\times4}^{(i)}(k_x,k_y,\theta)+H_{4\times4}^{(a)}(k_x,k_y,\theta).
\end{equation}
The isotropic part $H_{4\times4}^{(i)}(k_x,k_y,\theta)$ is written as~\cite{HOTIsm11,HOTIsm12}:
\begin{equation}
\label{eq:1DH16}
H_{4\times4}^{(i)}(k_x,k_y,\theta)=\begin{pmatrix}
\epsilon_{E1}(k_x,k_y) & -A_{1}k_{+} & R_{1}^{(i)}k_{-}^2 & S_{0}k_{-}\\
-A_{1}k_{-} & \epsilon_{H1}(k_x,k_y) & 0 & R_{2}^{(i)}k_{-}^2\\
R_{1}^{(i)}k_{+}^2 & 0 & \epsilon_{H2}(k_x,k_y)  & A_{2}k_{+}\\
S_{0}k_{+} & R_{2}^{(i)}k_{+}^2 & A_{2}k_{-} & \epsilon_{E2}(k_x,k_y) \end{pmatrix},
\end{equation}
where
\begin{eqnarray}
\label{eq:1DH17}
\epsilon_{E1}(k_x,k_y)=C_1+M_1-(D_{1}+B_{1})(k_x^2+k_y^2),\nonumber\\
\epsilon_{H1}(k_x,k_y)=C_1-M_1-(D_{1}-B_{1})(k_x^2+k_y^2),\nonumber\\
\epsilon_{E2}(k_x,k_y)=C_2+M_2-(D_{2}+B_{2})(k_x^2+k_y^2),\nonumber\\
\epsilon_{H2}(k_x,k_y)=C_2-M_2-(D_{2}-B_{2})(k_x^2+k_y^2),\nonumber\\
C_2=C_1+\dfrac{\Delta_{E1E2}-\Delta_{H1H2}}{2}.~~~~~~~~
\end{eqnarray}
In Eqs~(\ref{eq:1DH16}) and (\ref{eq:1DH17}), $C_{1,2}$, $M_{1,2}$, $A_{1,2}$, $B_{1,2}$, $D_{1,2}$, $S_0$ and $R_{1,2}^{(i)}$ are the structure parameters, which depend on $\theta$ (the growth direction of the QW), the QW profile and external conditions (such as temperature or hydrostatic pressure); $\Delta_{E1E2}$ and $\Delta_{H1H2}$ are the gaps between the \emph{E}1 and \emph{E}2 subbands and the \emph{H}1 and \emph{H}2 subbands, respectively.

The anisotropic term $H_{4\times4}^{(a)}(k_x,k_y,\theta)$ due to cubic symmetry of $H^{(a)}_L$ (see Eq.~(\ref{eq:1DH7})) has the form
\begin{multline}
\label{eq:1DH18}
H_{4\times4}^{(a)}(k_x,k_y,\theta)=\begin{pmatrix}
0 & 0 & -R_{1}^{(a)}k_{+}^2 & 0\\
0 & 0 & 0 & -R_{2}^{(a)}k_{+}^2\\
-R_{1}^{(a)}k_{-}^2 & 0 & 0  & 0\\
0 & -R_{2}^{(a)}k_{-}^2 & 0 & 0 \end{pmatrix}+ \\
+\sin^{2}{2\theta}\begin{pmatrix}
0 & 0 & -R_{1}^{(a)}k_{y}^2 & 0\\
0 & 0 & 0 & -R_{2}^{(a)}k_{y}^2\\
-R_{1}^{(a)}k_{y}^2 & 0 & 0  & 0\\
0 & -R_{2}^{(a)}k_{y}^2 & 0 & 0 \end{pmatrix}+\\
+\sin{2\theta}\begin{pmatrix}
0 & 0 & \tilde{R}_{1}^{(a)}(\theta) & 0\\
0 & 0 & 0 & \tilde{R}_{2}^{(a)}(\theta)\\
\tilde{R}_{1}^{(a)}(\theta) & 0 & 0  & 0\\
0 & \tilde{R}_{2}^{(a)}(\theta) & 0 & 0 \end{pmatrix}.
\end{multline}
In Eqs.~(\ref{eq:1DH16}) and (\ref{eq:1DH18}), $R_{1,2}^{(i)}$, $R_{1,2}^{(a)}$ and $\tilde{R}_{1,2}^{(a)}(\theta)$ are defined as
\begin{equation*}
\tilde{R}_{1,2}^{(a)}(\theta)=\int\limits_{-\infty}^{+\infty}f_{4,3}^{(E1,H1)}\left[
\left(k_{z}\dfrac{\sqrt{3}\hbar^2\left(\gamma_3(z)-\gamma_2(z)\right)}{m_0}k_{z}-\left(\sqrt{3}b-d\right)\left(\epsilon_{xx}-\epsilon_{zz}\right)\right)\dfrac{\sin{2\theta}}{4}
+\left(\sqrt{3}b-d\right)\epsilon_{yz}\dfrac{\cos{2\theta}}{2}\right]f_{6,5}^{(H2,E2)}dz,
\end{equation*}
\begin{eqnarray}
\label{eq:1DH19}
R_{1,2}^{(i)}=\dfrac{\hbar^2}{2m_0}\dfrac{\sqrt{3}}{2}\int\limits_{-\infty}^{+\infty}f_{4,3}^{(E1,H1)}\left(\gamma_3(z)+\gamma_2(z)\right)f_{6,5}^{(H2,E2)}dz,\nonumber\\
R_{1,2}^{(a)}=\dfrac{\hbar^2}{2m_0}\dfrac{\sqrt{3}}{2}\int\limits_{-\infty}^{+\infty}f_{4,3}^{(E1,H1)}\left(\gamma_3(z)-\gamma_2(z)\right)f_{6,5}^{(H2,E2)}dz.
\end{eqnarray}
Note that $\gamma_2(z)$ and $\gamma_3(z)$ are the symmetric functions of $z$ in accordance with the assumption made above. One should recall that the strain tensor components $\epsilon_{\alpha\beta}$ also depend on $\theta$; their expressions for the ($0mn$)-oriented QWs can be found in the Supplemental materials of Ref.~\cite{HOTIsm10}.

Up to now, it has been assumed that $x$ and $y$ axis are oriented along (100) and ($0n\bar{m}$) crystallographic directions, respectively. To this end, we consider the edge in an arbitrary direction $x'$, which has the angle $\varphi$ relative to the $x$ axis. To write the Hamiltonian in another coordinate system, where the $z'$ and $z$ axis coincide with [$0mn$] crystallographic orientation, one should rotate the electron momentum according to the transformation:
\begin{equation}
\label{eq:1Dedge10a}
\begin{pmatrix}
k_x \\
k_y
\end{pmatrix}=
\begin{pmatrix}
\cos{\varphi} & -\sin{\varphi} \\
\sin{\varphi} & \cos{\varphi}
\end{pmatrix}\begin{pmatrix}
k'_x \\
k'_y
\end{pmatrix}.
\end{equation}
Simultaneously with the transition from $k'_x$ and $k'_y$ to $k_x$ and $k_y$, one should also
apply a unitary transformation to the Hamiltonian (\ref{eq:1DH15}):
\begin{equation}
\label{eq:1Dedge10b}
H'_{4\times4}({k'}_x,{k'}_y,\theta,\varphi)=U(\varphi)H_{4\times4}(k_x,k_y,\theta)U(\varphi)^{-1},
\end{equation}
where
\begin{equation}
\label{eq:1Dedge10c}
U(\varphi)=\begin{pmatrix}
1 & 0 & 0 & 0 \\
0 & e^{i\varphi} & 0 & 0 \\
0 & 0 & e^{-2i\varphi} & 0 \\
0 & 0 & 0 & e^{-i\varphi} \end{pmatrix}.
\end{equation}

As it has been expected, ${H'}_{4\times4}^{(i)}(k'_x,k'_y,\theta)$ has the same form as ${H}_{4\times4}^{(i)}(k_x,k_y,\theta)$ in Eq.~(\ref{eq:1DH16}), while ${H'}_{4\times4}^{(a)}({k'}_x,{k'}_y,\theta,\varphi)$ becomes
\begin{multline}
\label{eq:1Dedge11}
{H'}_{4\times4}^{(a)}({k'}_x,{k'}_y,\theta,\varphi)=\begin{pmatrix}
0 & 0 & -R_{1}^{(a)}e^{i4\varphi}{k'}_{+}^2 & 0\\
0 & 0 & 0 & -R_{2}^{(a)}e^{i4\varphi}{k'}_{+}^2\\
-R_{1}^{(a)}e^{-i4\varphi}{k'}_{-}^2 & 0 & 0  & 0\\
0 & -R_{2}^{(a)}e^{-i4\varphi}{k'}_{-}^2 & 0 & 0 \end{pmatrix}+ \\
+\left({k'}_{y}\cos{\varphi}+{k'}_{x}\sin{\varphi}\right)^{2}\sin^{2}{2\theta}\begin{pmatrix}
0 & 0 & -R_{1}^{(a)}e^{i2\varphi} & 0\\
0 & 0 & 0 & -R_{2}^{(a)}e^{i2\varphi}\\
-R_{1}^{(a)}e^{-i2\varphi} & 0 & 0  & 0\\
0 & -R_{2}^{(a)}e^{-i2\varphi} & 0 & 0 \end{pmatrix}+ \\
+\sin{2\theta}\begin{pmatrix}
0 & 0 & \tilde{R}_{1}^{(a)}(\theta)e^{i2\varphi} & 0\\
0 & 0 & 0 & \tilde{R}_{2}^{(a)}(\theta)e^{i2\varphi}\\
\tilde{R}_{1}^{(a)}(\theta)e^{-i2\varphi} & 0 & 0  & 0\\
0 & \tilde{R}_{2}^{(a)}(\theta)e^{-i2\varphi} & 0 & 0 \end{pmatrix}.
\end{multline}
As expected from symmetry considerations, ${H'}_{4\times4}^{(a)}({k'}_x,{k'}_y,\theta,\varphi)$ in Eq.~(\ref{eq:1Dedge11}) for (001)- and (010)-oriented QWs features a $\pi/2$-periodicity, i.e., ${H'}_{4\times4}^{(a)}({k'}_x,{k'}_y,0,\varphi\pm{\pi/2})={H'}_{4\times4}^{(a)}({k'}_x,{k'}_y,0,\varphi)$ and ${H'}_{4\times4}^{(a)}({k'}_x,{k'}_y,\pi/2,\varphi\pm{\pi/2})={H'}_{4\times4}^{(a)}({k'}_x,{k'}_y,\pi/2,\varphi)$. This is clear from Eqs~(\ref{eq:1DH19}), which show that $R_{1,2}^{(0)}(0)=R_{1,2}^{(0)}(\pm{\pi/2})=0$.

Further, we omit the prime marks keeping in mind that orientation of new $x$ and $y$ axis does not coincide with the crystallographic directions in the most general case. Parameters involved in
$H_{4\times4}^{(a)}(k_x,k_y,\theta)$ and ${H}_{4\times4}^{(a)}({k}_x,{k}_y,\theta,\varphi)$ for the semiconductor QWs considered in the main text are provided in Table~\ref{tab:2}.

\subsection{Low-energy 1D Hamiltonian for the edge states in the (0$mn$) QWs}
To analyze the corner states in the QWs with double band inversion, we analytically derive the effective 1D Hamiltonian for the edge states. First, we split $H_{2D}(k_x,k_y,\theta,\varphi)$ (also see Eq.~(\ref{eq:1DH14})) into two terms so that the first term represent two independent BHZ-like models for the pairs of \emph{E}1-\emph{H}1 subbands and \emph{E}2-\emph{H}2 subbands, while the second term includes the inter-pairs mixing. Then, assuming the open-boundary conditions in a semi-infinite plane $y>0$, we solve the eigenvalue problems for independent BHZ-like blocks with $M_1<0$ and $M_2<0$ to find four wave functions at zero-wave vector along the boundary. Finally, we construct low-energy 1D Hamiltonian by projecting $H_{2D}(k_x,k_y,\theta,\varphi)$ onto the obtained set of the basis functions.


The edge wave functions (at zero wave-vector along the boundary) for two independent BHZ-like blocks for \emph{E}1-\emph{H}1 and \emph{E}2-\emph{H}2 subbands  in $H_{2D}(k_x,k_y,\theta,\varphi)$ are written as:
\begin{eqnarray}
\label{eq:1Dedge1}
|1,+\rangle=\dfrac{g_1(y)}{\sqrt{1+\eta_1^2}}\begin{pmatrix}
1 \\
\eta_1 \\
0 \\
0 \\
0 \\
0 \\
0 \\
0
\end{pmatrix},~~
|2,-\rangle=\dfrac{g_2(y)}{\sqrt{1+\eta_2^2}}\begin{pmatrix}
0 \\
0 \\
\eta_2 \\
1 \\
0 \\
0 \\
0 \\
0
\end{pmatrix},~~
|1,-\rangle=\dfrac{g_1(y)}{\sqrt{1+\eta_1^2}}\begin{pmatrix}
0 \\
0 \\
0 \\
0 \\
1 \\
\eta_1 \\
0 \\
0
\end{pmatrix},~~
|2,+\rangle=\dfrac{g_2(y)}{\sqrt{1+\eta_2^2}}\begin{pmatrix}
0 \\
0 \\
0 \\
0 \\
0 \\
0 \\
\eta_2 \\
1
\end{pmatrix},
\end{eqnarray}
where
\begin{eqnarray}
\label{eq:1Dedge2}
\eta_n^2=\dfrac{B_n+D_n}{B_n-D_n},~~~~~~~~~~~~~~~~~~~~~~\nonumber\\
g_n(y)=N_n\left(\lambda_n^{(I)},\lambda_n^{(II)}\right)\left\{e^{-\lambda_n^{(I)}y}-e^{-\lambda_n^{(II)}y}\right\},\nonumber\\
N_n\left(\lambda_n^{(I)},\lambda_n^{(II)}\right)=\sqrt{\left|2\lambda_n^{(I)}\lambda_n^{(II)}\dfrac{\lambda_n^{(I)}+\lambda_n^{(II)}}{\left(\lambda_n^{(I)}-\lambda_n^{(II)}\right)^2}\right|}.
\end{eqnarray}
Here, $n=1$ and $2$ correspond to the pairs of \emph{E}1-\emph{H}1 and \emph{E}2-\emph{H}2 subbands, respectively. In Eqs~(\ref{eq:1Dedge2}), while $\lambda_n^{(I)}$ and $\lambda_n^{(II)}$ have the form
\begin{eqnarray}
\label{eq:1Dedge2b}
\lambda_n^{(I,II)}=\sqrt{F_n\pm\sqrt{F_n^2-\dfrac{M_n^2}{B_n^2}}},\nonumber\\
F_n=\dfrac{A_n^2}{2(B_n^2-D_n^2)}-\dfrac{M_n}{B_n}.
\end{eqnarray}
It is clear that for the existence of the edge states, $\lambda_n^{(I)}$ and $\lambda_n^{(II)}$ should have a non-zero real part. This can be achieved if $\lambda_n^{(I,II)}$ are both real (at $F_n^2\geq{M_n^2}/{B_n^2}$). In the opposite case when ${M_n^2}/{B_n^2}>F_n^2$, the square of $\lambda_n^{(I,II)}$ are complex conjugated:
\begin{equation*}
\left\{\lambda_n^{(I,II)}\right\}^2=F_n{\pm}i\sqrt{\dfrac{M_n^2}{B_n^2}-F_n^2}.
\end{equation*}
To go further, it is convenient to present the square root from complex number $Q+iW$ in algebraic form $\sqrt{Q+iW}=\pm\left(a+ib\right)$, where
\begin{eqnarray*}
a=\sqrt{\dfrac{\sqrt{Q^2+W^2}+Q}{2}},~~~~~~\nonumber\\ b=\mathrm{sgn}(W)\sqrt{\dfrac{\sqrt{Q^2+W^2}-Q}{2}}.~~
\end{eqnarray*}
This allows presenting $\lambda_n^{(I)}$ and $\lambda_n^{(II)}$ in the form $\lambda_n^{(I,II)}={a_n}\pm{ib_n}$, where
\begin{eqnarray*}
a_n=\dfrac{\sqrt{2}}{2}\sqrt{\dfrac{M_n}{B_n}+F_n},~~\nonumber\\ b_n=\dfrac{\sqrt{2}}{2}\sqrt{\dfrac{M_n}{B_n}-F_n}.~~
\end{eqnarray*}
For the latter case, $g_n(y)$ in Eq.~(\ref{eq:1Dedge2}) can be written as
\begin{equation}
\label{eq:1Dedge2b}
g_n(y)=2\dfrac{\sqrt{|a_n|(a_n^2+b_n^2)}}{|b_n|}e^{-{a_n}y}\sin({b_n}y),
\end{equation}
where $a_n$ and $b_n$ are both real being defined as $\lambda_n^{(I,II)}={a_n}\pm{b_n}$.

Since we have set $k_y=-i\partial/\partial{y}$ to obtain Eqs.~(\ref{eq:1Dedge1}) and (\ref{eq:1Dedge2}), we also need to introduce the following matrix elements:
\begin{eqnarray}
\label{eq:1Dedge3}
\left\langle{k_y^2}\right\rangle_{nm}=\int\limits_{0}^{+\infty}g_n(y)\left(-\dfrac{\partial^2}{\partial{y^2}}\right)g_m(y)dy,~\nonumber\\
\left\langle{k_y}\right\rangle_{nm}=\int\limits_{0}^{+\infty}g_n(y)\left(-i\dfrac{\partial}{\partial{y}}\right)g_m(y)dy.
\end{eqnarray}
A straightforward calculation results in
\begin{eqnarray}
\label{eq:1Dedge4}
\left\langle{k_y^2}\right\rangle_{nm}=\left\langle{k_y^2}\right\rangle_{mn}=F(n,m)
\left(\lambda_n^{(I)}\lambda_n^{(II)}\lambda_m^{(I)}+\lambda_n^{(I)}\lambda_m^{(I)}\lambda_m^{(II)}+\lambda_n^{(I)}\lambda_n^{(II)}\lambda_m^{(II)}+\lambda_n^{(II)}\lambda_m^{(I)}\lambda_m^{(II)}\right),\nonumber\\
\left\langle{k_y}\right\rangle_{nm}=-\left\langle{k_y}\right\rangle_{mn}=-iF(n,m)
\left(\lambda_n^{(I)}\lambda_n^{(II)}-\lambda_m^{(I)}\lambda_m^{(II)}\right),~~~~~~~~~~~~~~~~~~~~~~~
\end{eqnarray}
where
\begin{equation}
\label{eq:1Dedge5}
F(n,m)=F(m,n)=N_nN_m\dfrac{\left(\lambda_n^{(I)}-\lambda_n^{(II)}\right)\left(\lambda_m^{(I)}-\lambda_m^{(II)}\right)} {\left(\lambda_n^{(I)}+\lambda_m^{(I)}\right)\left(\lambda_n^{(I)}+\lambda_m^{(II)}\right) \left(\lambda_n^{(II)}+\lambda_m^{(II)}\right)\left(\lambda_n^{(II)}+\lambda_m^{(I)}\right)}
\end{equation}
One can see that $\left\langle{k_y}\right\rangle_{nn}=0$ and $\left\langle{k_y^2}\right\rangle_{nn}=\lambda_n^{(I)}\lambda_n^{(II)}$.

Before going further, we note that Eqs~(\ref{eq:1Dedge4}) and (\ref{eq:1Dedge5}) are also valid for complex conjugated $\lambda_n^{(I,II)}$, i.e. for $\lambda_n^{(I,II)}={a_n}\pm{b_n}$. For the latter case, Eqs~(\ref{eq:1Dedge4}) can be rewritten as
\begin{eqnarray*}
\left\langle{k_y^2}\right\rangle_{nm}=\left\langle{k_y^2}\right\rangle_{mn}=8\dfrac{\sqrt{|a_n|(a_n^2+b_n^2)}\sqrt{|a_m|(a_m^2+b_m^2)}
\left(a_{n}^{2}a_{m}+a_{n}a_{m}^{2}+a_{n}b_{m}^{2}+a_{m}b_{n}^{2}\right)}
{\left[(a_{n}+a_{m})^2+(b_{n}-b_{m})^2\right]\left[(a_{n}+a_{m})^2+(b_{n}+b_{m})^2\right]},~~~~~~~~~~~~~~~~~~\nonumber\\~~~~~~~~~~~~~~~~~~~~
\left\langle{k_y}\right\rangle_{nm}=-\left\langle{k_y}\right\rangle_{mn}=-4i
\dfrac{\sqrt{|a_n|(a_n^2+b_n^2)}\sqrt{|a_m|(a_m^2+b_m^2)}
\left(a_{n}^{2}-a_{m}^{2}+b_{n}^{2}-b_{m}^{2}\right)}
{\left[(a_{n}+a_{m})^2+(b_{n}-b_{m})^2\right]\left[(a_{n}+a_{m})^2+(b_{n}+b_{m})^2\right]}.~~~~~~~~~~~~~~~~~~~~~~~
\end{eqnarray*}

Since projection of the blocks $H_{4\times4}(k_x,k_y,\theta,\varphi)$ and $H_{4\times4}^{*}(-k_x,-k_y,\theta,\varphi)$ onto the edge basis functions in Eq.~(\ref{eq:1Dedge1}) is performed independently, we further focus on the upper block $H_{4\times4}(k_x,k_y,\theta,\varphi)$ only. The procedure for the lower block $H_{4\times4}^{*}(-k_x,-k_y,\theta,\varphi)$ can be done in a similar manner. Thus, to project $H_{4\times4}(k_x,k_y)$, one has to consider only the states $|1,+\rangle$ and $|2,-\rangle$, which are reduced to
\begin{equation}
\label{eq:1Dedge6}
|1\rangle=\dfrac{g_1(y)}{\sqrt{1+\eta_1^2}}\begin{pmatrix}
1 \\
\eta_1 \\
0 \\
0
\end{pmatrix},~~~~~~~~~
|2\rangle=\dfrac{g_2(y)}{\sqrt{1+\eta_2^2}}\begin{pmatrix}
0 \\
0 \\
\eta_2 \\
1
\end{pmatrix},
\end{equation}
where the signs $\pm$ are omitted.

As mentioned above, for the projection of $H_{4\times4}(k_x,k_y,\theta,\varphi)$ onto the basis states $|1\rangle$ and $|2\rangle$, it is also convenient to present $H_{4\times4}(k_x,k_y,\theta,\varphi)$ in the form (cf. Eqs~(\ref{eq:1DH16}) and (\ref{eq:1Dedge11})):
\begin{equation}
\label{eq:1Dedge7}
H_{4\times4}(k_x,k_y,\theta,\varphi)=H_{2{\times}\mathrm{BHZ}}^{(i)}(k_x,k_y,\theta)+\widetilde{H}_{4\times4}^{(a)}(k_x,k_y,\theta,\varphi),
\end{equation}
where
\begin{equation}
\label{eq:1Dedge8}
H_{2{\times}\mathrm{BHZ}}^{(i)}(k_x,k_y,\theta)=\begin{pmatrix}
\epsilon_{E1}(k_x,k_y) & -A_{1}k_{+} & 0 & 0\\
-A_{1}k_{-} & \epsilon_{H1}(k_x,k_y) & 0 & 0\\
0 & 0 & \epsilon_{H2}(k_x,k_y)  & A_{2}k_{+}\\
0 & 0 & A_{2}k_{-} & \epsilon_{E2}(k_x,k_y)\end{pmatrix},
\end{equation}
and
\begin{multline}
\label{eq:1Dedge9}
\widetilde{H}_{4\times4}^{(a)}(k_x,k_y,\theta,\varphi)=\begin{pmatrix}
0 & 0 & R_{1}^{(i)}k_{-}^2 & S_{0}k_{-}\\
0 & 0 & 0 & R_{2}^{(i)}k_{-}^2\\
R_{1}^{(i)}k_{+}^2 & 0 & 0 & 0\\
S_{0}k_{+} & R_{2}^{(i)}k_{+}^2 & 0 & 0 \end{pmatrix}+ \\
+\begin{pmatrix}
0 & 0 & -R_{1}^{(a)}e^{i4\varphi}{k}_{+}^2 & 0\\
0 & 0 & 0 & -R_{2}^{(a)}e^{i4\varphi}{k}_{+}^2\\
-R_{1}^{(a)}e^{-i4\varphi}{k}_{-}^2 & 0 & 0  & 0\\
0 & -R_{2}^{(a)}e^{-i4\varphi}{k}_{-}^2 & 0 & 0 \end{pmatrix}+ \\
+\left({k}_{y}\cos{\varphi}+{k}_{x}\sin{\varphi}\right)^{2}\sin^{2}{2\theta}\begin{pmatrix}
0 & 0 & -R_{1}^{(a)}e^{i2\varphi} & 0\\
0 & 0 & 0 & -R_{2}^{(a)}e^{i2\varphi}\\
-R_{1}^{(a)}e^{-i2\varphi} & 0 & 0  & 0\\
0 & -R_{2}^{(a)}e^{-i2\varphi} & 0 & 0 \end{pmatrix}+ \\
+\sin{2\theta}\begin{pmatrix}
0 & 0 & \tilde{R}_{1}^{(a)}(\theta)e^{i2\varphi} & 0\\
0 & 0 & 0 & \tilde{R}_{2}^{(a)}(\theta)e^{i2\varphi}\\
\tilde{R}_{1}^{(a)}(\theta)e^{-i2\varphi} & 0 & 0  & 0\\
0 & \tilde{R}_{2}^{(a)}(\theta)e^{-i2\varphi} & 0 & 0 \end{pmatrix}.
\end{multline}

By using Eqs.~(\ref{eq:1Dedge2})--(\ref{eq:1Dedge5}) for integration along the $y$ axis, the projection of $H_{2{\times}\mathrm{BHZ}}^{(i)}(k_x,k_y,\theta)$ leads to
\begin{equation}
\label{eq:1Dedge12}
H_{1D}^{(i)}(k_{x},\theta)=\begin{pmatrix}
C_1-\dfrac{M_1D_1}{B_1}-\dfrac{2A_1\eta_1}{1+\eta_1^2}k_x &  0 \\
0 & C_2-\dfrac{M_2D_2}{B_2}+\dfrac{2A_2\eta_2}{1+\eta_2^2}k_x
\end{pmatrix}.
\end{equation}
Two blocks of $H_{1D}^{(i)}(k_{x},\theta)$ corresponds to the edge states resulting from inversion of the subband pairs $|E1,{+}\rangle$--$|H1,{+}\rangle$ and $|E2,{-}\rangle$--$|H2,{-}\rangle$ in the absence of their mixing. Therefore, the energies of these edge states cross at $k_x=k_c$:
\begin{equation}
\label{eq:1Dedge13}
k_c=\dfrac{C_1-C_2+\dfrac{M_2D_2}{B_2}-\dfrac{M_1D_1}{B_1}}{\dfrac{2A_1\eta_1}{1+\eta_1^2}+\dfrac{2A_2\eta_2}{1+\eta_2^2}}.
\end{equation}
Note that the crossing for other Kramer's partners occur at $k_x=-k_c$.

As clear from Eq.~(\ref{eq:1Dedge12}), the energy bands of $H_{1D}^{(i)}(k_{x},\theta)$ are nothing but a tilted 1D Dirac cone. In this case, the projection of $\widetilde{H}_{4\times4}^{(a)}(k_x,k_y,\theta,\varphi)$ in Eq.~(\ref{eq:1Dedge9}) results in anti-diagonal mass terms describing the band-gap opening. After straightforward calculation, the matrix element $\langle{1}|\widetilde{H}_{4\times4}^{(a)}(k_x,k_y,\theta,\varphi)|{2}\rangle$ is written as
\begin{multline}
\label{eq:1Dedge14}
\langle{1}|\widetilde{H}_{4\times4}^{(a)}|{2}\rangle= \left(\left\langle{k_y^2}\right\rangle_{12}-k_{x}^2-2k_x\left\langle{k_y}\right\rangle_{12}i\right)e^{i4\varphi}F_{a}
-\left(\left\langle{k_y^2}\right\rangle_{12}-k_{x}^2+2k_x\left\langle{k_y}\right\rangle_{12}i\right)F_{i}+
\left(k_{x}-\left\langle{k_y}\right\rangle_{12}i\right)F_{0}+\\
+e^{i2\varphi}\tilde{F}_{a}(\theta)\sin{2\theta}-\left(\left\langle{k_y^2}\right\rangle_{12}\cos^{2}{\varphi}+{k}_{x}^{2}\sin^{2}{\varphi}+k_x\left\langle{k_y}\right\rangle_{12}\sin{2\varphi}\right)
e^{i2\varphi}F_{a}\sin^{2}{2\theta}
\end{multline}
where
\begin{eqnarray}
\label{eq:1Dedge15}
F_{i}=\dfrac{R_{1}^{(i)}\eta_2+R_{2}^{(i)}\eta_1}{\sqrt{1+\eta_1^2}\sqrt{1+\eta_2^2}},~~~~~\nonumber\\
F_{a}=\dfrac{R_{1}^{(a)}\eta_2+R_{2}^{(a)}\eta_1}{\sqrt{1+\eta_1^2}\sqrt{1+\eta_2^2}},~~~~~\nonumber\\
\tilde{F}_{a}(\theta)=\dfrac{\tilde{R}_{1}^{(a)}(\theta)\eta_2+\tilde{R}_{2}^{(a)}(\theta)\eta_1}{\sqrt{1+\eta_1^2}\sqrt{1+\eta_2^2}},\nonumber\\
F_{0}=\dfrac{S_0}{\sqrt{1+\eta_1^2}\sqrt{1+\eta_2^2}}.~~~~~
\end{eqnarray}
The calculation of $\langle{2}|\widetilde{H}_{4\times4}^{(a)}|{1}\rangle$ is performed in the same way.

On the basis of Eq.~(\ref{eq:1Dedge4}), the matrix elements of $k_y$ can be presented in more convenient form
\begin{eqnarray}
\label{eq:1Dedge16}
\left\langle{k_y^2}\right\rangle_{12}=\left\langle{k_y^2}\right\rangle_{21}=\kappa_2,~~\nonumber\\
\left\langle{k_y}\right\rangle_{12}=-\left\langle{k_y}\right\rangle_{21}=-i\kappa_1.
\end{eqnarray}
The latter allows writing projection of $\widetilde{H}_{4\times4}^{(a)}(k_x,k_y,\theta,\varphi)$ in the form
\begin{multline}
\label{eq:1Dedge17}
H_{1D}^{(a)}(k_{x},\theta,\varphi)=\left[\left(F_i-F_a\cos{4\varphi}\right)k_{x}^2- \left(2F_i{\kappa_1}+2F_a{\kappa_1}\cos{4\varphi}-F_0\right)k_{x}+ F_a{\kappa_2}\cos{4\varphi}-F_i{\kappa_2}-F_0\kappa_1\right]\sigma_x+ \\
+\left[\tilde{F}_{a}(\theta)\cos{2\varphi}\sin{2\theta}
-F_{a}\sin^{2}{2\theta}\left(k_{x}^{2}\cos{2\varphi}\sin^{2}{\varphi}+
\kappa_{1}k_{x}\sin^{2}{2\varphi}+{\kappa_2}\cos{2\varphi}\cos^{2}{\varphi}\right)\right]\sigma_x+ \\
+F_a\sin{4\varphi}\left[k_x^2+2\kappa_1{k_x}-\kappa_2\right]\sigma_y+\\
\left[-\tilde{F}_{a}(\theta)\sin{2\varphi}\sin{2\theta}
+F_{a}\sin{2\varphi}\sin^{2}{2\theta}
\left(k_{x}^{2}\sin^{2}{\varphi}-\kappa_{1}k_{x}\cos{2\varphi}+{\kappa_2}\cos^{2}{\varphi}\right)\right]\sigma_y.
\end{multline}
Expanding now $H_{1D}^{(i)}(k_{x},\theta)+H_{1D}^{(a)}(k_{x},\theta,\varphi)$ around $\delta{k}=k_x-k_c$, we finally obtain the low-energy effective Hamiltonian for the tilted gapped 1D fermions:
\begin{equation}
\label{eq:1Dedge18}
H_{\mathrm{1D}}(\delta{k},\theta,\varphi)=\varepsilon_0+v_0\delta{k}\mathbf{I}_2+v_z\delta{k}\sigma_z+\left(m_y+v_y\delta{k}+\delta_y\delta{k}^2\right)\sigma_y+ \left(m_x+v_x\delta{k}+\delta_x\delta{k}^2\right)\sigma_x,
\end{equation}
where $\varepsilon_0$ is a constant corresponding to the energy of the crossing point at $k_x=k_c$ in the absence of $H_{1D}^{(a)}(k_{x},\theta,\varphi)$,
\begin{eqnarray*}
v_0=\dfrac{A_1\eta_1}{1+\eta_1^2}-\dfrac{A_2\eta_2}{1+\eta_2^2},\nonumber\\
v_z=\dfrac{A_1\eta_1}{1+\eta_1^2}+\dfrac{A_2\eta_2}{1+\eta_2^2},
\end{eqnarray*}
\begin{multline*}
m_x=\left(F_i-F_a\cos{4\varphi}\right)k_{c}^2+ \left(F_0-2{\kappa_1}(F_i+F_a\cos{4\varphi})\right)k_{c}+ F_a{\kappa_2}\cos{4\varphi}-F_i{\kappa_2}-F_0\kappa_1+\\
+\tilde{F}_{a}(\theta)\cos{2\varphi}\sin{2\theta}
-F_{a}\sin^{2}{2\theta}\left(k_{c}^{2}\cos{2\varphi}\sin^{2}{\varphi}+
\kappa_{1}k_{c}\sin^{2}{2\varphi}+{\kappa_2}\cos{2\varphi}\cos^{2}{\varphi}\right),
\end{multline*}
\begin{eqnarray}
\label{eq:1Dedge19}
m_y=F_a\sin{4\varphi}\left[k_c^2+2\kappa_1{k_c}-\kappa_2\right]
-\tilde{F}_{a}(\theta)\sin{2\varphi}\sin{2\theta}
+F_{a}\sin{2\varphi}\sin^{2}{2\theta}
\left(k_{c}^{2}\sin^{2}{\varphi}-\kappa_{1}k_{c}\cos{2\varphi}+{\kappa_2}\cos^{2}{\varphi}\right),\nonumber\\
v_x=F_{0}+2k_c\left(F_{i}-F_{a}\cos{4\varphi}\right)-2\kappa_1\left(F_{i}+F_{a}\cos{4\varphi}\right)
-F_{a}\sin^{2}{2\theta}\left(2k_{c}\cos{2\varphi}\sin^{2}{\varphi}+\kappa_{1}\sin^{2}{2\varphi}\right),\nonumber\\
v_y=2F_a\sin{4\varphi}\left(\kappa_1+k_c\right)+F_{a}\sin{2\varphi}\sin^{2}{2\theta}\left(2k_{c}\sin^{2}{\varphi}-\kappa_{1}\cos{2\varphi}\right),~~~~~~~~~~~~~\nonumber\\
\delta_x=F_i-F_a\cos{4\varphi}-F_{a}\cos{2\varphi}\sin^{2}{\varphi}\sin^{2}{2\theta},~~~~~~~~~~~~~~~~~~~~~~~~~~~~~~~\nonumber\\
\delta_y=F_a\sin{4\varphi}+F_{a}\sin{2\varphi}\sin^{2}{\varphi}\sin^{2}{2\theta}.~~~~~~~~~~~~~~~~~~~~~~~~~~~~~~~~~~~~
\end{eqnarray}
The analogous calculations for the block $H_{4\times4}^{*}(-k_x,-k_y,\theta,\varphi)$ results in
$H_{\mathrm{1D}}^{*}(-k_x-k_c,\theta,\varphi)$ (cf. Eq.~(\ref{eq:1Dedge18})). The parameters of the effective 1D edge Hamiltonian $H_{\mathrm{1D}}(\delta{k},\theta,\varphi)$ as a function of the edge orientation $\varphi$ for the three-layer InAs/GaInSb and double HgTe/CdHgTe QWs considered in the main text are provided in Fig.~\ref{Fig:SM1}.

\subsection{Energy of 0D corner states}
To calculate the energy of the corner states, we apply linear approximation for 1D edge $4\times4$ Hamiltonian consisting in two diagonal $2\times2$ blocks:
\begin{eqnarray}
\label{eq:SHcorn1new}
\tilde{H}_{\mathrm{1D}}^{(+)}(k,\theta,\varphi)=\varepsilon_0+v_0{k}\mathbf{I}_2+v_z{k}\sigma_z+(m_y+v_y{k})\sigma_y+ (m_x+v_x{k})\sigma_x,~\nonumber\\
\tilde{H}_{\mathrm{1D}}^{(-)}(\tilde{k},\theta,\varphi)=\varepsilon_0-v_0\tilde{k}\mathbf{I}_2-v_z\tilde{k}\sigma_z-(m_y-v_y\tilde{k})\sigma_y+ (m_x-v_x\tilde{k})\sigma_x,~
\end{eqnarray}
where $k=k_x-k_c$, $\tilde{k}=k_x+k_c$, $\tilde{H}_{\mathrm{1D}}^{(+)}(k,\theta,\varphi)$ and $\tilde{H}_{\mathrm{1D}}^{(-)}(\tilde{k},\theta,\varphi)$ are nothing but linearized $H_{\mathrm{1D}}(k_x-k_c,\theta,\varphi)$ and $H_{\mathrm{1D}}^{*}(-k_x-k_c,\theta,\varphi)$, respectively (see Sec.~B). Further, the constant $\varepsilon_0$ is omitted, while the eigenvalues of the 1D edge Hamiltonian are assumed to be counted from $\varepsilon_0$.

\begin{figure}[h!]
\includegraphics [width=1.0\columnwidth, keepaspectratio] {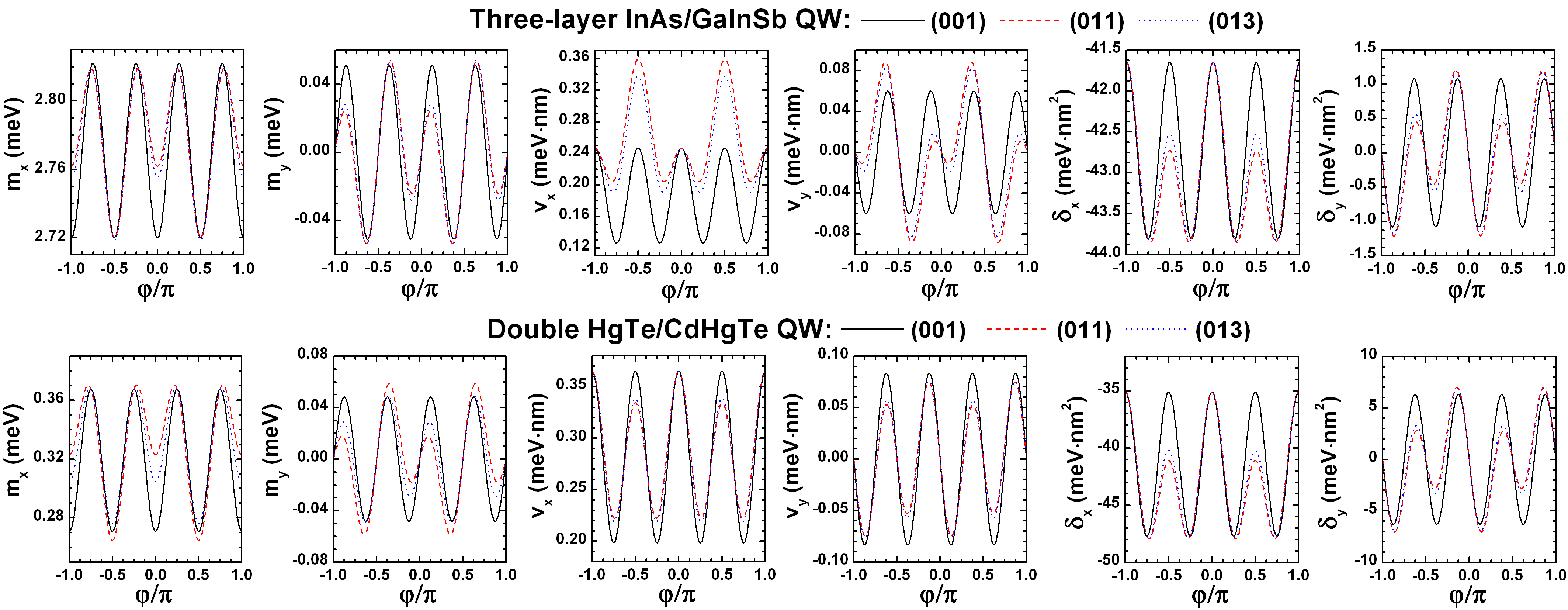} 
\caption{\label{Fig:SM1} Parameters of effective 1D edge Hamiltonian as a function of the edge orientation $\varphi$ for the three-layer InAs/GaInSb and double HgTe/CdHgTe QWs considered in the main text at different growth orientations. The edge independent parameters ($v_0$, $v_z$) equal to ($36.6$, $68.1$)~meV$\cdot$nm and ($-16.9$, $200.3$)~meV$\cdot$nm for the three-layer InAs/GaInSb and double HgTe/CdHgTe QWs, respectively.}
\end{figure}

First, one should make a certain remark significantly simplifying the calculations. With the parameters given in Table~\ref{tab:2}, it is clear that both $v_x$ and $v_y$ are significantly lower than $v_0$ and $v_z$ for any orientation of the edges (see Fig.~\ref{Fig:SM1}). The straightforward calculations shows that the presence of $v_x$ and $v_y$ in Eq.~(\ref{eq:SHcorn1new}) does not contribute significantly into the dispersion of the edge states as compared with other terms. Thus, one can neglect these terms in the first approximation and take them into account by using the perturbation theory.

Further, we focus on the upper block $\tilde{H}_{\mathrm{1D}}^{(+)}(k,\theta,\varphi)$, while the calculations for $\tilde{H}_{\mathrm{1D}}^{(-)}(\tilde{k},\theta,\varphi)$ are performed in a similar way. Let us now make a unitary transformation of $\tilde{H}_{\mathrm{1D}}^{(+)}(k,\theta,\varphi)$ as follows $H_{\mathrm{1D}}^{(+)}(k,\theta,\varphi)=U_{+}\tilde{H}_{\mathrm{1D}}^{(+)}(k,\theta,\varphi)U_{+}^{\dag}$, where
\begin{equation}
\label{eq:SHcorn1}
U_{+}=\dfrac{1}{\sqrt{2}}\begin{pmatrix}
1 &  i \\
-i & -1
\end{pmatrix}.
\end{equation}
The straightforward calculations results in
\begin{equation}
\label{eq:SHcorn2}
H_{\mathrm{1D}}^{(+)}(k,\theta,\varphi)
=v_{0}k\mathbf{I}_2-v_{z}k\sigma_y-m_y\sigma_z-m_x\sigma_x.
\end{equation}
It is clear that $H_{\mathrm{1D}}^{(+)}(k,\theta,\varphi)$ represent a 1D Dirac Hamiltonian, modified by "tilted" term $v_{0}k\mathbf{I}_2$ and additional mass term $m_x\sigma_x$.

Now, for a quantitative description, we define the coordinate $x$ along the curved edge so that $x=0$ corresponds to the meeting corner. In this case, $m_x$, $m_y$ in Eq.~(\ref{eq:SHcorn2}) are the function of $x$, and $k\equiv\hat{k}=-i\partial/\partial{x}$. Under this assumption, $H_{\mathrm{1D}}^{(+)}(\hat{k},\theta,\varphi)$ is defined in disjoint regions far from $x=0$. To define the 1D system fully, one needs to specify the boundary conditions that the wave functions must satisfy in the vicinity of $x=0$ in order to ensure that probability current along the curved edge is conserved. The current conservation implies that
\begin{equation}
\label{eq:SHcorn2new}
\Phi_1^{\dag}(v_{0}\mathbf{I}_2-v_{z}\sigma_y)\Phi_1=\Phi_2^{\dag}(v_{0}\mathbf{I}_2-v_{z}\sigma_y)\Phi_2,
\end{equation}
where $\Phi_1$ and $\Phi_2$ are the wave-functions defined from different sides of the corner. Note that specific type of the corner has not yet been determined.

Let us now discuss the general \emph{linear} boundary condition between $\Phi_1$ and $\Phi_2$. Let us assume that
\begin{equation}
\label{eq:SHcorn3new}
\Phi_1\big|_{x=-\eta}=\Pi\Phi_2\big|_{x=+\eta},
\end{equation}
where $\eta$ is a positive quantity and $\Pi$ is a unitary $2\times2$ matrix. Then Eq.~(\ref{eq:SHcorn2new}) will be satisfied if
$\Pi^{\dag}(v_{0}\mathbf{I}_2-v_{z}\sigma_y)\Pi=v_{0}\mathbf{I}_2-v_{z}\sigma_y$. The latter results in
\begin{equation}
\label{eq:SHcorn4new}
\Pi=\exp\left\{-i(\beta\sigma_y+\gamma\mathbf{I}_2)\right\},
\end{equation}
where $\beta$ and $\gamma$ are real parameters. Note that changing $\beta\rightarrow\beta+\pi$ and $\gamma\rightarrow\gamma+\pi$ has no effect on any physical quantities since this is just equivalent to changing $\Phi_1\rightarrow-\Phi_1$. Thus, one can assume that $\beta$ and $\gamma$ lie in the range from $-\pi/2$ to $\pi/2$.

The parameters $\beta$ and $\gamma$ in Eq.~(\ref{eq:SHcorn4new}) can be given a precise physical interpretation. Let us consider an additional $\delta$-function potential barrier in $H_{\mathrm{1D}}^{(+)}(k,\theta,\varphi)$ placed at $x=0$ given by $V_0\delta(0)$, where $V_0$ is a real parameter. Then, by integrating the Schr\"{o}dinger equation with the Hamiltonian in Eq.~(\ref{eq:SHcorn2}) through this potential, one can show that the wave function has indeed a discontinuity given by
\begin{equation}
\label{eq:SHcorn4newX}
\Phi_1\big|_{x=-\eta}=e^{-iV_0\left(\dfrac{v_z}{v_z^2-v_0^2}\sigma_y+\dfrac{v_0}{v_z^2-v_0^2}\mathbf{I}_2\right)}\Phi_2\big|_{x=+\eta},
\end{equation}
which coincides with Eqs~(\ref{eq:SHcorn3new}) and (\ref{eq:SHcorn4new}) if one defines $\beta$ and $\gamma$ as
\begin{equation}
\label{eq:SHcorn4newX2}
\beta=\dfrac{v_z}{v_z^2-v_0^2}V_0,~~~~~~~~~~~~\gamma=\dfrac{v_0}{v_z^2-v_0^2}V_0.
\end{equation}
The discontinuity is not surprising. We recall that for conventional non-relativistic Schr\"{o}dinger equation, which is second order in spatial derivatives, a $\delta$-function potential barrier leads to a discontinuity in the first derivative of the wave function. For the Dirac-like Hamiltonian, which is first order in spatial derivative, a $\delta$-function potential leads to a discontinuity in the wave function. Thus, Eqs~(\ref{eq:SHcorn3new}) and (\ref{eq:SHcorn4new}) at non-zero $\beta$ and $\gamma$ include the effects of a thin ($\delta$-like) barrier, which could possibly be present at the corner.

Representing $\beta$ and $\gamma$ as $\beta=\beta_2-\beta_1$ and $\gamma=\gamma_2-\gamma_1$, Eq.~(\ref{eq:SHcorn3new}) can be written in the form
\begin{equation}
\label{eq:SHcorn5new}
\Pi(\beta_1,\gamma_1)\Phi_1\big|_{x=-\eta}=\Pi(\beta_2,\gamma_2)\Phi_2\big|_{x=+\eta}.
\end{equation}
Thus, by means of Eq.~(\ref{eq:SHcorn4new}), one can write a new Hamiltonian $H_{\mathrm{1D}}^{(\mathrm{new})}(k,\tilde{\beta},\tilde{\gamma},\theta,\varphi)=\Pi(\tilde{\beta},\tilde{\gamma})H_{\mathrm{1D}}^{(+)}(k,\theta,\varphi)\Pi^{\dag}(\tilde{\beta},\tilde{\gamma})$ for the wave-functions $\Psi_{\mathrm{0D}}(x)=\Pi(\tilde{\beta},\tilde{\gamma})\Phi$ that are continuous in the vicinity of $x=0$:
\begin{equation}
\label{eq:SHcorn6new}
H_{\mathrm{1D}}^{(\mathrm{new})}(k,\tilde{\beta},\theta,\tilde{\gamma},\varphi)=v_{0}\hat{k}\mathbf{I}_2-v_{z}\hat{k}\sigma_y+M_z(x)\sigma_z-M_x(x)\sigma_x,
\end{equation}
where $M_z(x)$ and $M_x(x)$ are defined as
\begin{eqnarray}
\label{eq:SHcorn3}
M_z(x)=m_x(x)\sin2\tilde{\beta}-m_y(x)\cos2\tilde{\beta},\nonumber\\
M_x(x)=m_x(x)\cos2\tilde{\beta}+m_y(x)\sin2\tilde{\beta}.
\end{eqnarray}
One can see that $H_{\mathrm{1D}}^{(\mathrm{new})}(k,\tilde{\beta},\tilde{\gamma},\theta,\varphi)$ is actually independent of $\tilde{\gamma}$. Therefore, $\gamma$ in Eqs.~(\ref{eq:SHcorn4new})--(\ref{eq:SHcorn5new}) can be set to zero, and the boundary conditions at the corner can be considered to be characterized only by $\tilde{\beta}$ dependent on $x$. This case is discussed in the main text.

In view of the above, the Schr\"{o}dinger equation for the corner states takes the form
\begin{equation}
\label{eq:SHcorn5}
\left(-v_{z}\hat{k}\sigma_y+M_z(x)\sigma_z-M_x(x)\sigma_x\right)\Psi_{\mathrm{0D}}(x)=\left(E-v_{0}\hat{k}\right)\mathbf{I}_2\Psi_{\mathrm{0D}}(x).
\end{equation}
Let us act by the matrix operator from the left-hand side of Eq.~(\ref{eq:SHcorn5}) on both sides of this equation. This leads to
\begin{equation}
\label{eq:SHcorn6}
\bigg\{\left(v_{z}^2\hat{k}^2+M_z^2+M_x^2-(E-v_{0}\hat{k})^2\right)\mathbf{I}_2+
v_{z}\begin{pmatrix}
-M_x' &  -M_z' \\
-M_z' & M_x'
\end{pmatrix}
-v_{0}\begin{pmatrix}
-iM_z' &  iM_x' \\
iM_x' & iM_z'
\end{pmatrix}\bigg\}\Psi_{\mathrm{0D}}(x)=0,
\end{equation}
where the prime denotes the derivative with respect to $x$.

To find an exact solution of Eq.~(\ref{eq:SHcorn6}), we further restrict ourselves to the case, in which $M_z(x)$ and $M_x(x)$ are all proportional to each other
\begin{equation}
\label{eq:SHcorn7}
M_x(x)=\alpha{M_z(x)}+m,
\end{equation}
where
\begin{eqnarray}
\label{eq:SHcorn8}
\alpha=\dfrac{M_x(-\infty)-M_x(+\infty)}{M_z(-\infty)-M_z(+\infty)},~~~~~~~~~~~~\nonumber\\
m=\dfrac{M_z(-\infty)M_x(+\infty)-M_z(+\infty)M_x(-\infty)}{M_z(-\infty)-M_z(+\infty)}.
\end{eqnarray}
We note that Eq.~(\ref{eq:SHcorn7}) is a good approximation only for the relatively sharp functions varying in the vicinity of $x=0$. It is clear that Eq.~(\ref{eq:SHcorn7}) becomes \emph{exact} in the limit of the step-like functions $M_x(x)$ and $M_z(x)$.
The latter corresponds to the corner shown in Fig.~3 in the main text. Thus, Eq.~(\ref{eq:SHcorn6}) reads
\begin{equation}
\label{eq:SHcorn9}
\bigg\{\left(v_{z}^2\hat{k}^2+M_z(x)^2+\left\{\alpha{M_z(x)}+m\right\}^2-(E-v_{0}\hat{k})^2\right)\mathbf{I}_2+
M_z'\begin{pmatrix}
-v_{z}\alpha+iv_{0} & -v_{z}-iv_{0}\alpha \\
-v_{z}-iv_{0}\alpha & v_{z}\alpha-iv_{0}
\end{pmatrix}\bigg\}\Psi_{\mathrm{0D}}(x)=0.
\end{equation}
Hence solutions of Eq.~(\ref{eq:SHcorn9}) may be constructed as follows:
\begin{equation}
\label{eq:SHcorn10}
\Psi_{\mathrm{0D}}(x)={\chi}\psi(x),
\end{equation}
where ${\chi}$ is the spin part of the wave function satisfying equation
\begin{equation*}
\begin{pmatrix}
-v_{z}\alpha+iv_{0} & -v_{z}-iv_{0}\alpha \\
-v_{z}-iv_{0}\alpha & v_{z}\alpha-iv_{0}
\end{pmatrix}{\chi}=\nu{\chi},
\end{equation*}
with eigenvalues $\nu=\pm\sqrt{1+\alpha^2}\sqrt{v_{z}^2-v_{0}^2}$.

The equation for the coordinate part $\psi(x)$ can be written as
\begin{equation}
\label{eq:SHcorn11}
\Bigg\{\left(\sqrt{v_{z}^2-v_{0}^2}\hat{k}+\dfrac{Ev_{0}}{\sqrt{v_{z}^2-v_{0}^2}}\right)^2
+\left(\sqrt{1+\alpha^2}M_z(x)+\dfrac{m\alpha}{\sqrt{1+\alpha^2}}\right)^2
-\dfrac{E^2v_{z}^2}{v_{z}^2-v_{0}^2}+\dfrac{m^2}{1+\alpha^2}
+{\nu}M_z'\Bigg\}\psi(x)=0.
\end{equation}
Finally, by introducing a new variable $\tilde{x}=x/\sqrt{v_{z}^2-v_{0}^2}$ and representing $\psi(x)$ in the form
\begin{equation}
\label{eq:SHcorn12}
\psi(x)=\tilde{\psi}(\tilde{x})e^{\displaystyle{-i\tilde{x}\frac{Ev_{0}}{\sqrt{v_{z}^2-v_{0}^2}}}},
\end{equation}
we arrive at the following equation:
\begin{equation}
\label{eq:SHcorn13}
\Bigg\{\hat{\tilde{k}}^2
+\tilde{W}(\tilde{x})^2+\sigma\tilde{W}(\tilde{x})'\Bigg\}\tilde{\psi}(\tilde{x})=\varepsilon\tilde{\psi}(\tilde{x}),
\end{equation}
where $\sigma=\pm{1}$ (the sign of $\sigma$ coincides with those for $\nu$), and $\varepsilon$ and $\tilde{W}(\tilde{x})$ are defined as
\begin{eqnarray}
\label{eq:SHcorn14}
\varepsilon=\dfrac{E^2v_{z}^2}{v_{z}^2-v_{0}^2}-\dfrac{m^2}{1+\alpha^2},~~~~~~~~\nonumber\\
\tilde{W}(\tilde{x})=\sqrt{1+\alpha^2}M_z+\dfrac{m\alpha}{\sqrt{1+\alpha^2}}.
\end{eqnarray}

As seen from Eq.~(\ref{eq:SHcorn13}), it is the common Schr\"{o}dinger equation with a specific potential, which is a linear combination of the square of the derivative of the same function $\tilde{W}(\tilde{x})$.
It possesses a special symmetry and represents the formulation of supersymmetric quantum mechanics~\cite{HOTIsm20}.
The supersymmetric potential $\tilde{W}(\tilde{x})$ allows for factorization of Eq.~(\ref{eq:SHcorn13}):
\begin{equation}
\label{eq:SHcorn15}
\left(-i\hat{\tilde{k}}-\sigma\tilde{W}(\tilde{x})\right)\left(i\hat{\tilde{k}}-\sigma\tilde{W}(\tilde{x})\right)\tilde{\psi}(\tilde{x})=\varepsilon\tilde{\psi}(\tilde{x}),
\end{equation}
If the signs of the asymptotics $\tilde{W}(+\infty)$ and $\tilde{W}(-\infty)$ are opposite, i.e.
\begin{equation}
\label{eq:SHcorn16}
\left(M_z(+\infty)+\dfrac{m\alpha}{1+\alpha^2}\right)\left(M_z(-\infty)+\dfrac{m\alpha}{1+\alpha^2}\right)<0,
\end{equation}
Eq.~(\ref{eq:SHcorn15}) always has a localized solution $\tilde{\psi}(\tilde{x})$ with $\varepsilon=0$,
which converts the second brackets into zero:
\begin{equation}
\label{eq:SHcorn17}
\left(\dfrac{d}{d\tilde{x}}-\sigma\tilde{W}(\tilde{x})\right)\tilde{\psi}(\tilde{x})=0.
\end{equation}
Solution of this equation has the form
\begin{equation}
\label{eq:SHcorn18}
\tilde{\psi}(\tilde{x})\sim{e}^{\displaystyle\sigma\int\limits_0^{\tilde{x}}\tilde{W}(z)dz},
\end{equation}
where the sign of $\sigma$ should be chosen in accordance with normalized condition of $\tilde{\psi}(\tilde{x})$.
If $\tilde{W}(+\infty)>0$, $\sigma=-1$, while for $\tilde{W}(+\infty)<0$, $\sigma=1$.
We must note that these two cases are not equivalent.
One could see that the values of $\sigma=1$ and $\sigma=-1$ correspond to the internal and external corners at the same positions of two edges.
Further, we show that the localized states for these two corners have different energies.

By using Eqs.~(\ref{eq:SHcorn10}), (\ref{eq:SHcorn12}), (\ref{eq:SHcorn14}) and (\ref{eq:SHcorn18}), the wave function of the corner state is expressed as follows:
\begin{equation}
\label{eq:SHcorn19}
\Psi_{\mathrm{0D}}(x)=C
\begin{pmatrix}
v_{z}\alpha-\sigma\sqrt{1+\alpha^2}\sqrt{v_{z}^2-v_{0}^2}-iv_{0} \\
v_{z}+iv_{0}\alpha
\end{pmatrix}
e^{\displaystyle{-ix\frac{Ev_{0}}{v_{z}^2-v_{0}^2}+\frac{\sigma}{\sqrt{1+\alpha^2}\sqrt{v_{z}^2-v_{0}^2}}\int\limits_0^{x}\left\{(1+\alpha^2)M_z(z)+m\alpha\right\}dz}},
\end{equation}
where $C$ is the normalization constant. Now substituting $\Psi_{\mathrm{0D}}(x)$ into Eq.~(\ref{eq:SHcorn5}), one can show that
\begin{equation}
\label{eq:SHcorn20}
E=\dfrac{{\sigma}m}{\sqrt{1+\alpha^2}}
\dfrac{\sqrt{v_{z}^2-v_{0}^2}}{v_z}.
\end{equation}
Thus, the localized states for the internal and external corners, corresponding to the same positions of two edges,
have opposite energies. One can verified that
\begin{equation*}
E^2<\dfrac{m^2}{1+\alpha^2}\leq{M_x(x)^2+M_z(x)^2}=m_x(x)^2+m_y(x)^2.
\end{equation*}

Let us make few remarks concerning the results obtained above. First, we have found a localized 0D corner state, whose energy does not depend on the specific type of functions $M_z(x)$ and $M_x(x)$.
The existence of such localized state is guaranteed by the two conditions defined by Eq.~(\ref{eq:SHcorn7}) and Eq.~(\ref{eq:SHcorn16}). The latter can be also written in equivalent form
\begin{equation}
\label{eq:SHcorn21}
\left(M_z(+\infty)+{\alpha}M_x(+\infty)\right)\left(M_z(-\infty)+{\alpha}M_x(-\infty)\right)<0.
\end{equation}

We now take into account the small terms previously neglected in Eq.~(\ref{eq:SHcorn1new}).
Note that before applying the perturbation theory on the basis of the wave-function given be Eq.~(\ref{eq:SHcorn19}), one should perform the unitary transformation to get the correction to $H_{\mathrm{1D}}^{(\mathrm{new})}(k,\tilde{\beta},\theta,\varphi)$:

\begin{equation}
\label{eq:SHcorn24}
\delta{H}_{\mathrm{1D}}^{(\mathrm{new})}(k,\tilde{\beta},\theta,\varphi)=
\left(v_x(x)\sin2\tilde{\beta}-v_y(x)\cos2\tilde{\beta}\right)k\sigma_z
-\left(v_x(x)\cos2\tilde{\beta}+v_y(x)\sin2\tilde{\beta}\right)k\sigma_x.
\end{equation}
The first-order correction to the energy of the localized corner state in Eq.~(\ref{eq:SHcorn20}) can be calculated analytically assuming that $M_x(x)$ and $M_z(x)$ are the step-like functions,
for which the theory above is \emph{exact}. The straightforward calculations on the basis of $\Psi_{\mathrm{0D}}(x)$ leads to the first-order energy shift:
\begin{equation}
\label{eq:SHcorn22}
\delta{E}=-\dfrac{v_{0}m}{v_{z}^2(1+\alpha^2)}\left(V(-\infty)\dfrac{\lambda(+\infty)}{\lambda(-\infty)+\lambda(\infty)}+V(+\infty)\dfrac{\lambda(-\infty)}{\lambda(-\infty)+\lambda(\infty)}\right),
\end{equation}
where
\begin{eqnarray}
\label{eq:SHcorn23}
V(x)=(v_x+\alpha{v_y})\cos2\tilde{\beta}+(v_y-\alpha{v_x})\sin2\tilde{\beta},\nonumber\\
\lambda(\pm\infty)=\dfrac{(1+\alpha^2)M_z(\pm\infty)+m\alpha}{\sqrt{1+\alpha^2}\sqrt{v_{z}^2-v_{0}^2}}.~~~~~~~~
\end{eqnarray}

%

\newpage

\begin{table}[h!]
\caption{\label{tab:1} Parities of the envelope functions of multi-band \textbf{k$\cdot$p} Hamiltonian at zero electron momentum in the QW plane.}
\begin{ruledtabular}
\begin{tabular}{c|cccccccc}
QW subband & \emph{E}\{2k+1\} & \emph{E}\{2k+2\} & \emph{H}\{2k+1\} & \emph{H}\{2k+2\} & \emph{LH}\{2k+1\} & \emph{LH}\{2k+2\} \\[6pt]
\hline
Even  & $f_{1,2}^{(E\{2k+1\})}(z)$ & $f_{4,5}^{(E\{2k+2\})}(z)$ & $f_{3,6}^{(H\{2k+1\})}(z)$ & -- & $f_{4,5}^{(LH\{2k+1\})}(z)$ & $f_{1,2}^{(LH\{2k+2\})}(z)$ \\[6pt]
Odd   & $f_{4,5}^{(E\{2k+1\})}(z)$ & $f_{1,2}^{(E\{2k+2\})}(z)$ & -- & $f_{3,6}^{(H\{2k+2\})}(z)$ & $f_{1,2}^{(LH\{2k+1\})}(z)$ & $f_{4,5}^{(LH\{2k+2\})}(z)$
\end{tabular}
\end{ruledtabular}
\end{table}

\begin{table}[h!]
\caption{\label{tab:2} Parameters involved in the effective 2D Hamiltonian for the three-layer InAs/GaInSb and double HgTe/CdHgTe QWs considered in the main text.}
\begin{ruledtabular}
\begin{tabular}{c|c|c|c|c|c|c|c|c|c}
2D system & \makecell{$C_1$ \\ (meV)} & \makecell{$C_2$ \\ (meV)} & \makecell{$M_1$ \\ (meV)} & \makecell{$M_2$ \\ (meV)} & \makecell{$B_1$ \\ (meV$\cdot$nm$^2$)} & \makecell{$B_2$ \\ (meV$\cdot$nm$^2$)} & \makecell{$D_1$ \\ (meV$\cdot$nm$^2$)} & \makecell{$D_2$ \\ (meV$\cdot$nm$^2$)} & \makecell{$S_0$ \\ (meV$\cdot$nm)} \\
\hline
Three-layer InAs/GaInSb QW & 51.60 & 34.93 & -45.35 & -20.78 & -720 & -520 & -51 & -320 & -37 \\
Double HgTe/CdHgTe QW & -37.90  & -27.43 & -13.99 & -3.52 & -1175 & -695 & -1025 & -545 & 2
\end{tabular}
~\\~

\begin{tabular}{c|c|c|c|c|c|c}
2D system & \makecell{$A_1$ \\ (meV$\cdot$nm)} & \makecell{$A_2$ \\ (meV$\cdot$nm)} & \makecell{$R_{1}^{(i)}$ \\ (meV$\cdot$nm$^2$)} & \makecell{$R_{2}^{(i)}$ \\ (meV$\cdot$nm$^2$)} & \makecell{$R_{1}^{(a)}$ \\ (meV$\cdot$nm$^2$)} & \makecell{$R_{2}^{(a)}$ \\ (meV$\cdot$nm$^2$)} \\
\hline
Three-layer InAs/GaInSb QW & 105 & 40 & -56 & -27 & -1.1 & -1.3  \\
Double HgTe/CdHgTe QW & 375 & 350 & -320 & 110 & -12.0 & -10.8
\end{tabular}
~\\~

\begin{tabular}{c|c|c|c|c|c|c}
2D system & \makecell{$\tilde{R}_{1}^{(a)}$~[001] \\ (meV$\cdot$nm)} & \makecell{$\tilde{R}_{2}^{(a)}$~[001] \\ (meV$\cdot$nm)} & \makecell{$\tilde{R}_{1}^{(a)}$~[011] \\ (meV$\cdot$nm)} &\makecell{$\tilde{R}_{2}^{(a)}$~[011] \\ (meV$\cdot$nm)} & \makecell{$\tilde{R}_{1}^{(a)}$~[013] \\ (meV$\cdot$nm)} & \makecell{$\tilde{R}_{2}^{(a)}$~[013] \\ (meV$\cdot$nm)} \\
\hline
Three-layer InAs/GaInSb QW & 0 & 0 & -3.6$\cdot10^{-3}$ & -2.8$\cdot10^{-4}$ & -1.6$\cdot10^{-3}$ & -2.1$\cdot10^{-4}$  \\
Double HgTe/CdHgTe QW & 0 & 0 & -2.4$\cdot10^{-2}$ & -1.6$\cdot10^{-3}$ & -1.4$\cdot10^{-2}$ & -1.2$\cdot10^{-3}$
\end{tabular}
\end{ruledtabular}
\end{table}
\end{document}